\documentclass[11pt]{article}
\usepackage{iclr2026_conference,times}
\usepackage{enumitem}

\usepackage{amsmath,amsfonts,bm}

\def\eqref#1{equation~\ref{#1}}

\def\1{\bm{1}}

\DeclareMathAlphabet{\mathsfit}{\encodingdefault}{\sfdefault}{m}{sl}
\SetMathAlphabet{\mathsfit}{bold}{\encodingdefault}{\sfdefault}{bx}{n}

\usepackage{makecell}
\usepackage{mathtools}
\usepackage{amssymb}
\usepackage{float}
\usepackage{caption}
\usepackage{graphicx}
\usepackage{wrapfig}
\usepackage{multirow}
\usepackage{pifont}
\usepackage{hyperref}
\usepackage{url}

\AtEndPreamble{
    \usepackage[capitalize]{cleveref}
    \crefname{section}{Sec.}{Secs.}
    \Crefname{section}{Section}{Sections}
    \crefname{table}{Tab.}{Tabs.}
    \Crefname{table}{Table}{Tables}
}

\usepackage{graphicx}
\usepackage{makecell}
\usepackage{xcolor}
\usepackage{xspace}
\usepackage{array}
\usepackage{wrapfig} 
\usepackage{colortbl}

\usepackage{booktabs}

\usepackage{hyperref}

\usepackage{tabularray}
\usepackage{multirow}
\usepackage{tcolorbox}
\usepackage[normalem]{ulem}
\usepackage{pdfpages}

\usepackage{subcaption} 

\newcommand{\mysec}[1]{\textit{\textbf{#1}}}

\newcommand{\tool}[0]{MULocBench\xspace}

\usepackage{hyperref}
\usepackage{url}

\title{A Benchmark for Localizing Code and Non-Code Issues in Software Projects} 

\author{
Zejun Zhang, Jian Wang, Qingyun Yang, Yifan Pan, Yi Tang, Yi Li, \\\textbf{\ Zhenchang Xing, Tian Zhang, Xuandong Li,  Guoan Zhang} 
}

\iclrfinalcopy

\begin{document}

\maketitle

\begin{abstract} 
Accurate project localization (e.g., files and functions) for issue resolution is a critical first step in software maintenance. 
However, existing benchmarks for issue localization, such as SWE-Bench and LocBench, are limited. 
They focus predominantly on pull-request issues and code locations, ignoring other evidence and non-code files such as commits, comments, configurations, and documentation. 
To address this gap, we introduce \tool, a comprehensive dataset of 1,100 issues from 46 popular GitHub Python projects.
Comparing with existing benchmarks, \tool offers greater diversity in issue types, root causes, location scopes, and file types, providing a more realistic testbed for evaluation. 
Using this benchmark, we assess the performance of state-of-the-art localization methods and five LLM-based prompting strategies. 
Our results reveal significant limitations in current techniques: even at the file level, performance metrics (Acc@5, F1) remain below 40\%.
This underscores the challenge of generalizing to realistic, multi-faceted issue resolution.
To enable future research on project localization for issue resolution, we publicly release \tool at https://huggingface.co/datasets/somethingone/\tool. 



\end{abstract}

\section{Introduction}
Modern software projects are inherently complex. 
They often consist of thousands of files spanning code, configurations, tests, and documentation. 
The complexity making developers routinely encounter a wide spectrum of issues, ranging from runtime failures and unexpected results to enhancement requests and usage questions. 
A prerequisite for resolving these issues is to accurately identify the locations, such as the relevant files and functions.

Existing benchmarks have advanced research on issue localization. 
SWE-Bench~\cite{jimenez2023swe} collects 2,294 issues with pull requests from 12 Python projects, primarily targeting bug fixing. 
To encourage adoption, it releases SWE-bench Lite, a subset of 300 instances. 
Most Recently, LocBench~\cite{chen-etal-2025-locagent} expands the scope to 560 issues with more diverse types. 
Despite these efforts, existing benchmarks remain limited. 
They focus narrowly on code and rely solely on pull requests as evidence of resolution, overlooking other resolution signals and artifact types. 
In practice, issues may be resolved not only through pull requests but also via commits and even explicit comments. 
Moreover, project locations often involve files beyond source code, such as configurations, documentation, and third-party libraries.

To better reflect these real-world scenarios, we present \tool, a dataset of 1,100 issues drawn from 46 popular GitHub Python projects. 
Each issue is linked to a pull request, commit, or resolution-confirming comment, and includes detailed location information—project name, file path, class name (if applicable), function name (if applicable), and line numbers (if applicable). 
\tool offers broader and more balanced coverage across issue types, root causes, location scopes, and file types, providing a richer foundation for evaluating localization approaches. 

To evaluate how well existing techniques support realistic issue localization, we conduct two experiments based on \tool. 
We first investigate the effectiveness of state-of-the-art approaches( retrieval-based, procedure-based, and agent-based methods) on locating identification within in-project Python code. 
LocAgent and OpenHands with Claude3.5 achieve the best and comparable results, with Acc@5 reaching at most 35.2\%, 28.5\%, and 22.0\% at the file, class, and function levels.
The result is substantially lower than the 60\%+ Acc@5 consistently observed across all three levels on SWE-Bench Lite and LocBench. 
We then further examine whether large language models (LLMs) can go further, identifying locations not only in code files and in-project files. 
We design five LLM-based prompting strategies with varying levels of contextual support, from closed-book to pipeline(Location-Hint guided) localization. 
The best performance with Location-Hint guided reaches only around 32.5\%, 14.5\%, and 13.1\% Acc@5 at the file, class, and function levels. 

In summary, the main contributions are: (1) \tool—a project location benchmark covering 1,100 issues with rich issues and locations. (2) A comprehensive comparison with state-of-art location benchmarks. (3) A systematic evaluation highlighting the limitations of current localization techniques and LLMs in project localization for issue resolution. 

\vspace{-3mm}
\section{Benchmark Construction}\label{method}

\begin{figure}
  \centering
    \setlength{\abovecaptionskip}{1mm}
    \includegraphics[width=5.4in]{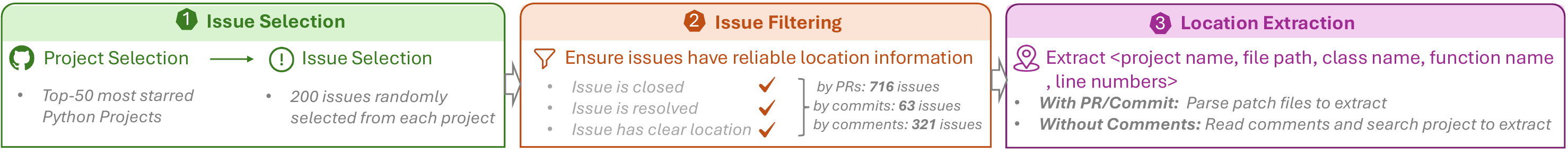}
     \caption{\tool Construction Overview}
    \label{fig:app_overview} 
    \vspace{-7mm} 
\end{figure}
Project developers encounter numerous and diverse issues in the process of software development. 
Location identification is a prerequisite for resolving software issues. 
To build a benchmark for location information, we draw on GitHub Issues. 
The benchmark construction is a three-stage pipleine, as shown in Figure~\ref{fig:app_overview}. 
First, we select the top-50 starred Python repositories and randomly sample 200 issues per project to balance diversity and manual effort. 
Then, we retain only closed, resolved issues with reliable location via merged PRs, commits, or explicit comments; for issues without PRs/commits, comments must clearly state the location. 
Finally, we extract locations (project name, file path, class, function, line) automatically from PR/commit patches or manually from resolution comments. 
The details are as follows:

\noindent \textbf{(1) Issue selection:} 
The process begins by specifying a project set, then selecting issues within those projects. 
We focus on the top 50 most-starred GitHub Python projects, which typically have large communities and rich issue discussions, making them ideal for constructing representative and diverse benchmarks. 
Python is chosen for its top-ranked popularity and broad domain applicability, which often leads to diverse project-related issues. 
For each project, we randomly sample 200 issues to balance manual effort with question diversity.

\noindent \textbf{(2) Issue filtering:} Reliable location information is critical for our study, but many issues lack such information due to being open, unresolved, or having ambiguous descriptions. 
To obtain reliable location data, we apply three filtering conditions: selecting closed issues, retaining those with confirmed resolutions, and keeping only issues with clear location information. 
After filtering,  there are 1,100 issues from 46 projects. 
Among them, 716 have pull requests, 63 have commits, and 321 have resolution-related comments. 
The details of filtering stage are in Appendix~\ref{issue_filter_process}. 

\noindent \textbf{(3) Location extraction:} Location information includes \textit{project name}, \textit{file path}, \textit{class name}, \textit{function name}, and \textit{line numbers}. 
For issues linked to pull requests or commits, location information is automatically extracted by parsing the corresponding patch files. 
For issues without linked pull requests or commits, we manually extract location information by analyzing the resolution-related comments. 
All manual extracted location have been confirmed by authors. 
For example, consider the comment: just changed line \textit{v2.VideoCapture(camera\_index)} in ui.py.  
By inspecting the repository, we identify the location as follows: the project name is \texttt{hacksider/Deep-Live-Cam}, the file path is \texttt{modules/ui.py}, the class name is empty (as the line is not within a class), the function name is \texttt{render\_video\_preview}, and the line number is 203. 
In another example, the location refers to a configuration file rather than source code.  
The comment states: ``It works with Python 3.10.12 when numpy version is changed to numpy==1.22.0 in requirements.txt.''  
Here, the file path is \texttt{requirements.txt}, the class name and function name are empty, and the line number is 4. 

\vspace{-3mm}
\section{Empirical Analysis}\label{empi_ana}

\begin{figure*}
  \centering
    \setlength{\abovecaptionskip}{2mm}
    \includegraphics[width=5.5in]{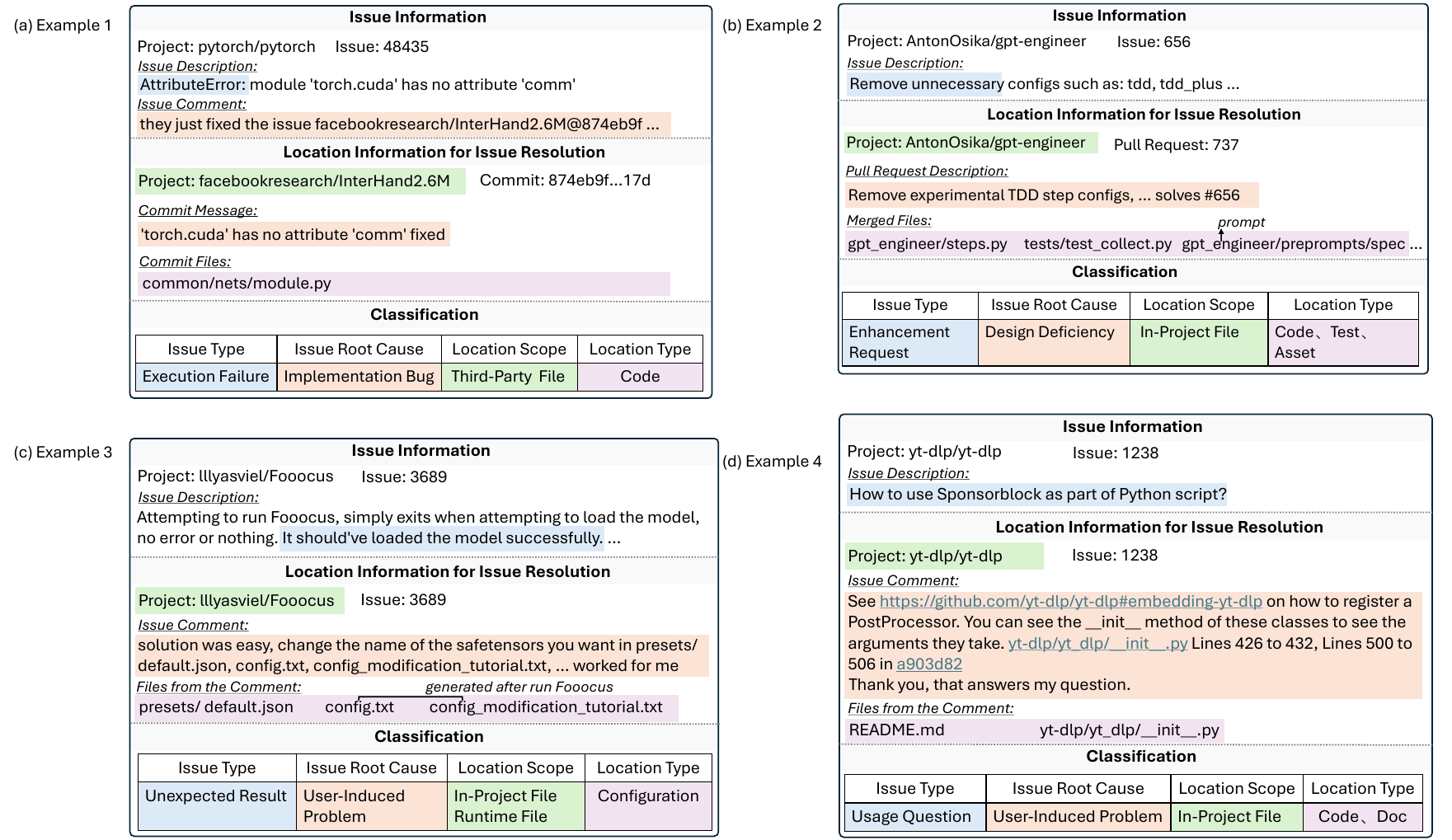}%
    \caption{Four issues illustrating different issue types, reasons, location scopes and types. 
}  
    \label{fig:comparison_iss_num} 
    \vspace{-0.6cm}
\end{figure*}

After constructing \tool, a project-level, location-centric benchmark, we conduct an empirical study of issue reports and their resolution locations.  
Prior work~\cite{jimenez2023swe,chen-etal-2025-locagent} has mainly curated issue–patch pairs and focused on bug fixing, but offered little analysis of issue types, causes, or affected files.  
Such analysis is crucial for understanding benchmark characteristics and enabling meaningful comparisons.  
Illustrative examples of issue types, root causes, location scopes, and location types are shown in Figure~\ref{fig:comparison_iss_num}. 
The analysis methods and definitions of issue types, reasons, and location scopes and types detailed in Appendix~\ref{analysis_method_iss_location} and Appendix~\ref{analysis_result_iss_location}.

\noindent \textbf{Issue Types} capture the problems developers raise in issue reports, such as execution errors or usage questions.  
We classify issues into four categories: Execution Failures (39.9\%), Unexpected Results (23.7\%), Enhancement Requests (25.1\%), and Usage Questions (11.3\%). 
For example, in panel (d) of Figure~\ref{fig:comparison_iss_num}, a developer asks how to use \texttt{Sponsorblock} as part of a Python script, which is classified as a Usage Question because it reflects a request for guidance.  


\noindent \textbf{Root Causes of Issues} capture why problems arise.  
We categorize them into Implementation Bugs (34.5\%), Design Deficiencies (36.3\%), and User-Induced Problems (29.2\%). 
For example, in panel (c) of Figure~\ref{fig:comparison_iss_num}, developers fail to load a model in \texttt{Fooocus} until they correct their own configuration settings. 
It represents a User-Induced Problem, as the issue stems from incorrect usage rather than flaws in the project itself.  



\noindent \textbf{Location Scopes} describe where fixes occur, including In-Project Files (94.1\%), Runtime Files (2.2\%), Third-Party Files (2.1\%), and User-Authored Files (3.0\%). 
The percentages exceed 100\% because an issue may span multiple scopes. 
For example, in panel (a) of Figure~\ref{fig:comparison_iss_num}, an issue reported in the \texttt{pytorch/pytorch} project is ultimately traced to an implementation bug in the external dependency \texttt{facebookresearch/InterHand2.6M}, which is classified as a Third-Party File because it belongs to an external project rather than the PyTorch repository itself.

\noindent \textbf{Location Types} describe file categories affected in resolution: Code (80.8\%), Test (23.5\%), Configuration (15.2\%), Documentation (23.5\%), and Asset (4.4\%). 
The total exceeds 100\% because an issue resolution may involve multiple file types. 
For example, in panel (b) of Figure~\ref{fig:comparison_iss_num}, the issue resolution location involves \texttt{gpt\_engineer/preprompts/spec}, which is classified as an Asset because it is prompt text used as input rather than executable code.  



\section{COMPARISON WITH STATE-OF-ART LOCATION BENCHMARK}\label{compare_benchmarks}
\begin{figure*}

  \centering
    \setlength{\abovecaptionskip}{1mm}
    \includegraphics[width=5.3in]{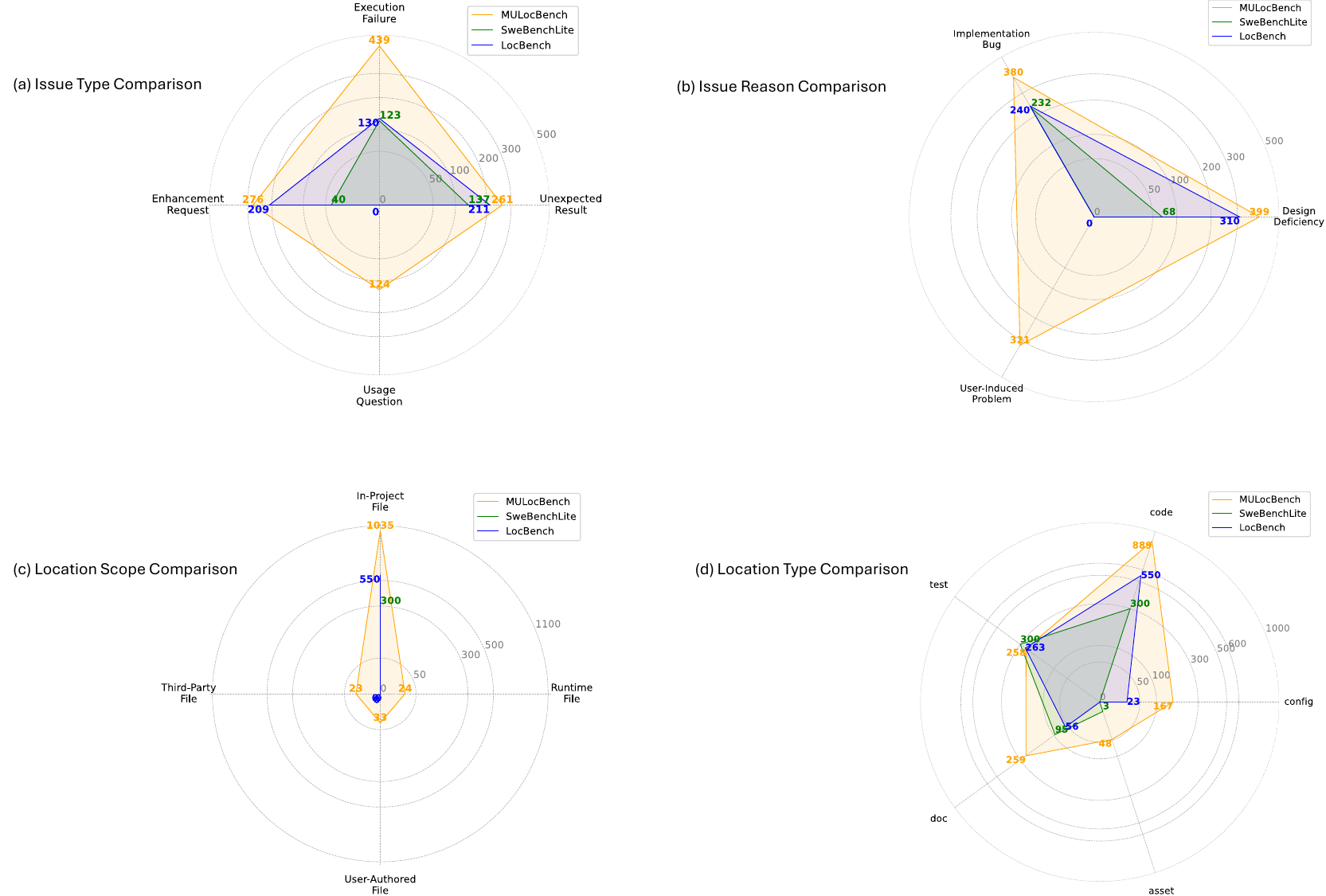}
    \caption[An Issue location can invovlve multiple location scopes and types.]{ Issue number comparison between \tool, SWE-Bench Lite and LocBench
    }
    \label{fig:comparison_benchmarks} 
    \vspace{-0.7cm}

\end{figure*}
\subsection{State-of-Art Location Benchmark from GitHub Issues} 

\textbf{SWE-Bench Lite}~\cite{jimenez2023swe} is a widely adopted benchmark that collects GitHub issues together with the corresponding code patches. It contains 300 issues from 12 Python projects. 
However, its collection only includes pull requests that modify Python code and the corresponding test files, which limits its ability to capture the broader context of real-world issue resolution.

\noindent \textbf{LocBench}~\cite{chen-etal-2025-locagent} is a new published benchmark that covers more diverse issues and a larger set of Python projects. 
Its latest release contains 560 issues, but we identified 10 cases where the pull requests do not correspond to the issues or are duplicated, leaving 550 valid issues from 163 repositories. 
Similar to SWE-Bench Lite, it only includes issues with pull requests that modify Python code, but not necessarily the associated test files. 

\noindent \textbf{\tool}. In contrast, our \tool is constructed by randomly sampling issues from the top-50 most-starred Python projects, resulting in 1,100 issues from 46 Python projects. 
It includes resolution evidence from pull requests, commits, and comments, and covers more diverse issue types and project locations to better reflect real-world development practices. 
\subsection{Location Benchmark Comparison for Issue Resolution} 

To better understand the differences among benchmarks, we conduct a systematic comparison of \tool, SWE-Bench Lite, and LocBench along four dimensions: issue types, root causes, location scopes, and location types. 
Figure~\ref{fig:comparison_benchmarks} shows the comparison results across these benchmarks.

\noindent \textbf{Issue Type:} SWE-Bench Lite mainly covers Execution Failures (123 issues) and Unexpected Results (137), with limited Enhancement Requests (40) and no Usage Questions (0). 
LocBench expands the scope by significantly increasing the number of Unexpected Results (211) and Enhancement Requests (209), while Execution Failures (130) remain similar and Usage Questions (0) are still absent. 
In contrast, \tool provides a comprehensive coverage across all four types, including Execution Failures (439), Unexpected Results (261), Enhancement Requests (276), and Usage Questions (124 issues). 

\noindent \textbf{Issue Cause:} SWE-Bench Lite is dominated by Implementation Bugs (232 issues), with limited Design Deficiencies (68) and no coverage of User Problems (0), implicitly assuming that issues mainly stem from project defects. 
LocBench shifts the distribution, reporting 310 issues in Design Deficiencies and 240 issues in Implementation Bugs, but still excludes User Problems (0 issues), thus overemphasizing design flaws while ignoring user-side challenges. 
In contrast, \tool achieves broader balance by covering 380 issues in Implementation Bugs, 399 issues in Design Deficiencies, and 321 issues in User Problems, recognizing that many issues also arise from user misunderstandings or misuse rather than project-side faults. 


\noindent \textbf{Location Scope:} SWE-Bench Lite (300 issues) and LocBench (550 issues) only include In-Project Files, ignoring all other contexts. 
In contrast, \tool not only contains a larger number of In-Project Files (1,035 issues) but also incorporates 80 issues beyond the project, including 23 in Third-Party Files, 24 in Runtime Files, and 33 in User-Authored Files. 
This broader coverage makes \tool better aligned with real-world maintenance scenarios, where issues can extend beyond the project's static code base. 

\noindent \textbf{Location Type:} Both SWE-Bench Lite and LocBench are largely code-centric, with all issues involving code files. 
SWE-Bench Lite also includes 300 issues in test files, while its coverage of other file types is limited: 95 in documentation, 3 in assets, and none in configuration files. 
LocBench shows a similar pattern, with 263 issues in test files, only 56 in documentation, 23 in configuration, and none in assets. 
In contrast, \tool offers a much richer distribution of location files: 889 issues in code files, 258 in test files, 259 in documentation, 167 in configuration, and 48 in assets. 
This diversity better reflects the multifaceted composition of real-world projects, where issue resolution often extends beyond source code to involve configurations, documentation, and supporting resources.

Overall, both SWE-Bench Lite and LocBench remain narrow in scope: they emphasize specific issue types, focus heavily on project-side defects, restrict locations to in-project files, and are largely code-centric in artifact coverage. 
In contrast, \tool achieves broader and more balanced coverage across issue types, causes, scopes, and file categories, offering a more faithful representation of real-world issue resolution scenarios. 

\section{Effectiveness of State-of-the-Art Approaches in Code Localization on PLocBench}\label{rq1}
With the popularity of SWE-Bench Lite, state-of-the-art approaches have been widely studied in the context of Python code localization.
To assess their generality, we evaluate these approaches on \tool using in-project Python code files, since they are limited to handling such files. 

\subsection{Experimental Settings}
\mysec{Large Language Models.}
We select two leading large language models, gpt-4o-mini-2024-07-18 (GPT-4o-mini)~\cite{gpt4o} and claude-3-5-haiku-20241022 (Claude3.5)~\cite{claude}. 
Both models support extended context windows, making them well-suited for our tasks that involve long queries and large software projects. 
Moreover, they are widely adopted and acknowledged by recent studies in software engineering, demonstrating strong reasoning capabilities and retrieval-augmented performance in similar scenarios~\cite{chen-etal-2025-locagent,xia2024agentless,jimenez2023swe,reddy2025swerank}. 
We set the temperature to 0 to ensure that the outputs are mostly deterministic. 

\mysec{Approaches.}
There are three main types of state-of-the-art localization approaches: 

\noindent (1) Retrieval-based approach: We adopt BM25~\cite{robertson2009probabilistic}, a robust and simple term-matching method widely used in software engineering tasks such as code search and issue localization~\cite{jimenez2023swe,chen-etal-2025-locagent}.  



\noindent (2) Procedure-based approach: Agentless~\cite{xia2024agentless} is the most effective procedure-based issue localization approach to date. 
It performs hierarchical localization using a combination of prompting-based and embedding-based retrieval empowered by LLMs. 

\noindent (3) Agent-based approach: We select two advanced and widely used agent-based approaches: LocAgent~\cite{chen-etal-2025-locagent} and OpenHands~\cite{wang2024openhands}.
LocAgent proposes a graph-oriented LLM-agent framework for issue localization and is currently the most effective agent-based localization method. 
OpenHands is a React-style agent framework that enables LLMs to invoke tools for iterative reasoning and action to resolve issues.  


\mysec{Metrics.}
Following prior state-of-the-art approaches~\cite{chen-etal-2025-locagent,xia2024agentless}, which treat issue localization as a ranking task, we evaluate performance using $Top@k Acc$, where $k$ is set to 1 and 5, reflecting practical tolerance levels in real-world settings.
This metric deems localization successful if all relevant code locations are correctly identified within the top-k results. 

To allow a more fine-grained assessment of localization quality, we also compute precision, recall, and F1-score. 
For an issue, a true positive occurs when a location given by an approach exists in the benchmark. 
A false positive occurs when a location given by an approach does not exist in the benchmark. 
A false negative occurs when a configuration not given by an approach exists in the benchmark. 
We denote the number of true positives, false positives and false
negatives as $TP$, $FP$ and $FN$. 
We calculate the precision as $P = \frac{TP}{TP+FP}$, the recall as $R = \frac{TP}{TP+FN}$, and the F1-score  as $F1 = \frac{2\times P \times R}{P+R}$. 

Similar to previous localization tasks, we evaluate all metrics at three location levels: file, class, and function. 
Consistent with previous studies, we exclude the line-level comparison due to its instability and limited practical value: this level is rarely assessed and models typically achieve a accuracy lower than 1\%. 
Moreover, file-, class-, or function-level localization is generally sufficient for developers to resolve issues, which aligns with how developers typically act on localization. 

\mysec{DataSet.} Since state-of-the-art approaches~\cite{chen-etal-2025-locagent,xia2024agentless,reddy2025swerank,jimenez2023swe} typically localize only in-project Python files, even though pull requests may also modify other files such as JavaScript, documentation, or configuration, we follow the same setting and retain only locations in in-project Python code files, resulting in 842 issues.

\begin{table*}
    \centering
    \small
    \resizebox{1\textwidth}{!}{

\begin{tabular}{ll ccccc ccccc ccccc}
        \toprule
\multirow{2}{*}{Type} & \multirow{2}{*}{LLM} & \multicolumn{5}{c}{File(\%)} & \multicolumn{5}{c}{Class(\%)} & \multicolumn{5}{c}{Function(\%)}\tabularnewline
        \cmidrule(lr){3-7} \cmidrule(lr){8-12} \cmidrule(lr){13-17}

 &  & Acc@1 & Acc@5 & P & R & F1 & Acc@1 & Acc@5 & P & R & F1 & Acc@1 & Acc@5 & P & R & F1\tabularnewline
\midrule 
\midrule
\multicolumn{2}{l}{BM25} & 9.4 & 9.4 & 16.4 & 11.5 & 13.5 & 7.0 & 8.0 & 12.1 & 8.2 & 9.8 & 2.9 & 4.6 & 5.6 & 5.9 & 5.8\tabularnewline
\midrule 
\multirow{3}{*}{Agentless} &  GPT-4o-mini & 10.2 & 18.6 & 15.1 & 13.1 & 14.1 & 5.8 & 9.5 & 11.8 & 7.8 & 9.4 & 4.1 & 8.7 & 5.1 & 8.0 & 6.3\tabularnewline
 & Claude-3.5 & 12.2 & 23.9 & 18.6 & 15.6 & 17.0 & 3.3 & 4.5 & 12.7 & 2.5 & 4.2 & 6.4 & 11.0 & 11.2 & 5.3 & 7.2\tabularnewline
\midrule 
\multirow{3}{*}{LocAgent} & GPT-4o-mini & 11.7 & 16.1 & 38.0 & 12.9 & 19.2 & 6.6 & 10.8 & 19.6 & 7.2 & 10.6 & 4.3 & 9.4 & 23.5 & 6.6 & 10.3\tabularnewline
 & Claude-3.5 & 28.5 & 35.2 & 49.6 & 30.9 & 38.1 & 17.6 & 25.9 & 30.8 & 18.9 & 23.5 & 14.9 & 21.9 & 29.5 & 13.5 & 18.5\tabularnewline
\midrule 
\multirow{3}{*}{OpenHands} & GPT-4o-mini & 14.0 & 24.1 & 39.2 & 16.7 & 23.5 & 11.6 & 16.1 & 22.8 & 11.8 & 15.6 & 7.7 & 12.8 & 17.1 & 8.1 & 11.0\tabularnewline
&Claude-3.5  & 24.1 & 33.5 & 49.5 & 26.2 & 34.3 & 21.3 & 28.5 & 36.2 & 20.4 & 26.1 & 11.8 & 22.0 & 23.8 & 13.8 & 17.5 \\
        \bottomrule

\end{tabular}
}
    \caption{
    Performance of state-of-the-art methods on Python code localization in \tool.
    }
    \vspace{-.5cm}
    \label{tab:rq1_acc_comparison}
\end{table*}

\subsection{Result} \mysec{Approach Comparison.} Table~\ref{tab:rq1_acc_comparison} shows the performance of the state-of-the-art methods on the Python-only subset of the \tool at three localization levels: file, class, and function. 
Each method is evaluated using Acc@1, Acc@5, precision, recall, and F1-score. 
Performance basically degrades as the required localization granularity becomes finer (File $>$ Class $>$ Function). 
The four approaches show different performance: 
\textbf{LocAgent and OpenHands} yield the best and comparable results, followed by Agentless and then
BM25. 
The best results only reach 35.2\% Acc@5 / 38.1\% F1 at the file level, 28.5\% Acc@5 / 26.1\% F1 at the class level, and 22.0\% Acc@5 / 18.5\% F1 at the function level. 
Notably, LocAgent and OpenHands with Claude3.5 achieves over 80\%, 70\%, 60\% Acc@5 at file, class and function levels on SWE-Bench Lite~\cite{chen-etal-2025-locagent}, but with limited performance on our \tool. 
This suggests that the limited generality of state-of-art approaches on issue localization. 
The analysis result of each approach is provided in Appendix~\ref{rq1_approach_comparison}

\mysec{Model Comparison.} \textbf{Claude 3.5 outperforms GPT-4o-mini} across all metrics in most LLM-based approaches and localization levels. 
It achieves gains of +2.0–16.8\% in Acc@1, +2.3–18.9\% in Acc@5, +0.9–13.4\% in precision, +1.5–18.0\% in recall, and +0.9–18.9\% in F1 across the file, class, and function levels. 
An exception occurs under the Agentless at the class level, where Claude 3.5 underperforms GPT-4o-mini by 2.5\% in Acc@1, 5.0\% in Acc@5, and 5.2\% in F1. 

\begin{figure*}
  \centering
    \setlength{\abovecaptionskip}{1mm}
    \includegraphics[width=5.5in]{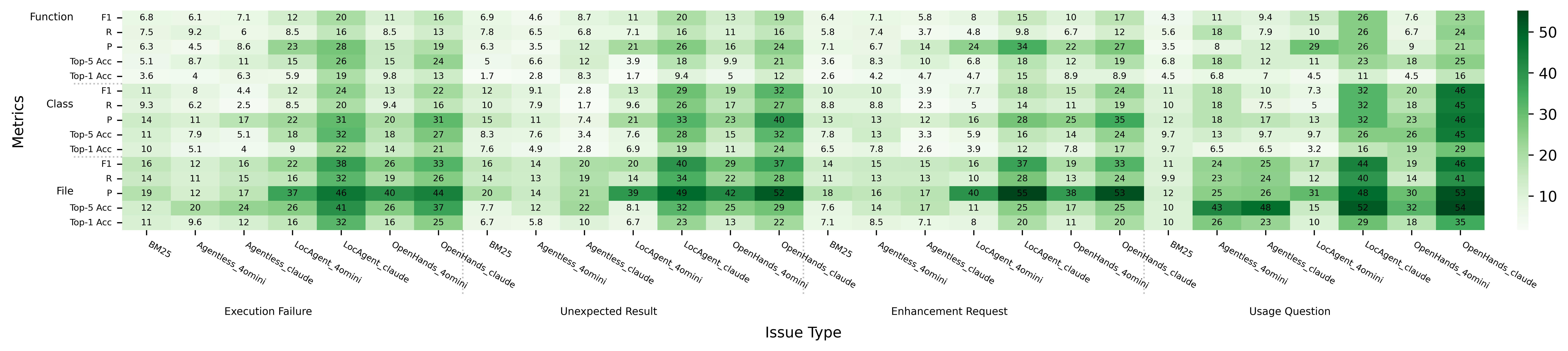}
    \caption{Performance comparison of different issue types on \tool with Python files. }  
    \label{fig:iss_type_rq1} 
    \vspace{-0.5cm}
\end{figure*}

\mysec{Performance Comparison across Issue Types.} Figure~\ref{fig:iss_type_rq1} shows the performance comparison of different issue types on \tool with only Python files. 
\textbf{Execution Failures} achieve moderate results, with LocAgent+Claude3.5 reaching up to 41\% Acc@5 and 38\% F1 at the file level, and their performance still competitive 26\%-32\% Acc@5 and 20\%-24\% F1 finer levels. 
\textbf{Unexpected Results and Enhancement Requests} yield the weakest outcomes, with the highest Acc@5 capped at around 30\%. 
Particularly, the precision is comparable to or even higher than that of other issue types. 
It indicates that while models often miss some relevant locations (low recall), the predicted locations have similar confidence to those of other issue types. 
\textbf{Usage Questions} achieve the strongest performance, with 54\% Acc@5 and 46\% F1 at the file level, likely because usage-related issues are described more concretely, providing clearer cues for localization.

\begin{figure*}
  \centering
    \setlength{\abovecaptionskip}{1mm}
    \includegraphics[width=5.5in]{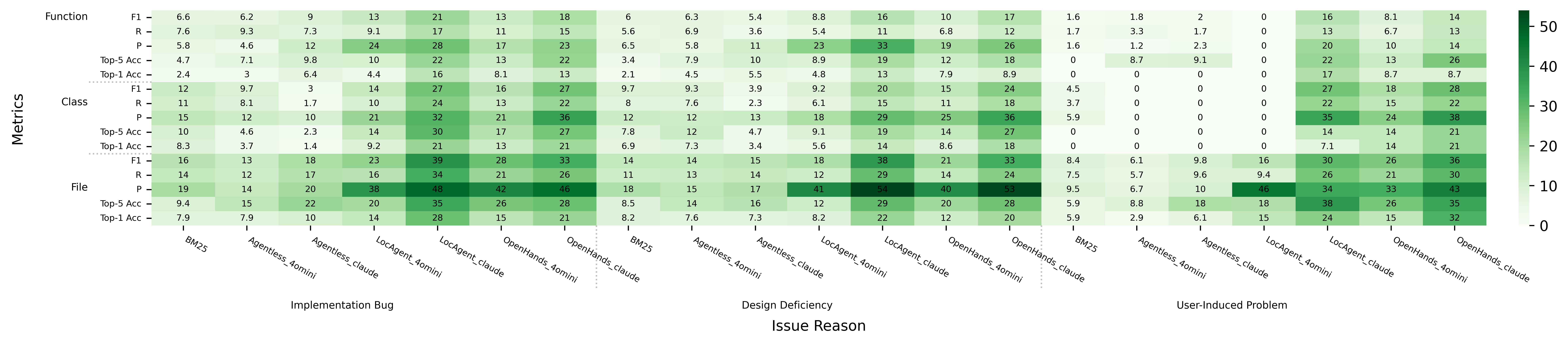}
    \caption{Performance comparison of different issue reasons on \tool with Python files. }  
    \label{fig:iss_reason_rq1} 
    \vspace{-0.5cm}
\end{figure*}

\mysec{Peformance Comparison across Issue Reasons.} Figure~\ref{fig:iss_reason_rq1} shows the performance comparison of different issue reasons on \tool with only Python files. 
\textbf{Implementation Bugs} show the most stable performance, with LocAgent+Claude 3.5 achieving the highest results: up to 35\% Acc@5 and 39\% F1 at the file level. 
Even at the class and function levels, performance remains relatively stable, with Acc@5 and F1 falling in the 22\%–30\%. 
This result reflects the fact that most existing datasets and methods in software engineering research are concentrated on bug-fixing scenarios. 
\textbf{Design Deficiencies} show the weakest results overall, with file-level Acc@5 around 29\%. 
However, the precision is relatively high (29\%-54\%) at three levels, comparable to or even better than other issue reasons. 
This aligns with the nature of enhancement request issues. 
\textbf{User-Induced Problems} show intermediate performance, with LocAgent + Claude 3.5 achieving the best file-level result of 38\% Acc@5. 
At the class and function levels, however, performance remains modest, with Acc@5 and F1 ranging from 14\% to 27\%.

\section{Effectiveness of LLMs in Project Location on Full \tool}\label{rq2}

State-of-the-art localization techniques are restricted to analyzing source code written in a single programming language (e.g., Python) and are constrained to the project repository. 
However, as reflected in the \tool benchmark, real-world issue localization is far more complex: relevant locations may appear in non-code files, span multiple file types, or even lie outside the scope of the project. 
Motivated by the success of LLMs, we investigate their effectiveness on the full \tool, which reflects realistic issue localization scenarios. 

\subsection{Experimental Settings} 
The evaluation metrics and LLMs used here are the same as in Section~\ref{rq1}. 

\mysec{LLM-based Approaches} To systematically explore the capabilities of LLMs in localization, we design five LLM-based approaches that represent different levels of information availability and reasoning complexity. 
These variants are inspired by common strategies in software engineering and recent successes of LLMs in code-related tasks~\cite{}. 
The prompts are shown in Appendix~\ref{prompt_template}

\noindent \textbf{(1) Closed-Book LLM:} In the closed-book setting, we directly prompt LLMs to generate localization predictions for GitHub issues by only providing Github issue report. 
This approach evaluates the model's ability to rely solely on its prior knowledge. 

\noindent \textbf{(2) Project-Structure LLM:} Directly feeding the entire project content into the LLM is computationally infeasible due to input length limitations. 
Given that our task is localization, we instead provide the model with project structure information to assist LLM' reasoning. 
The project structure includes a hierarchical representation of the directory tree, along with file names, file extensions, and their relative paths. 

\noindent  \textbf{(3) Location-Hint LLM:} Software projects often contain hundreds or thousands of files, which makes localization difficult due to the large and noisy search space.
We give the LLM location scope (e.g., within-project, third-party, user-authored files) and location type (e.g., code, configuration, documentation) as hints. 
We also provide the project structure but filter out files that do not match the specified location type, so the model can search within a smaller and more relevant subset.

\noindent  \textbf{(4) Pipeline (Project-Structure Guided) LLM:} 
A common way to apply LLM to complex tasks is to split the process into sequential steps. 
Following this idea, we design a two-stage pipeline for issue localization. 
In the first stage, we provide the issue report together with the project structure to let the LLM predict the candidate files. 
In the second stage, we give the predicted files along with their class and function names to the LLM to identify finer-grained elements such as the specific class, functions, and line numbers. 
We deliberately avoid supplying full file content since it often exceeds context length, slows inference, and yields limited additional benefit. 

\noindent  \textbf{(5) Pipeline (Location-Hint Guided) LLM:} 
Similar to the structure-guided pipeline, this approach decomposes issue localization into two stages.
In the first stage, we provide the issue report together with the location scope and location type as hints, and restrict the project structure to only include files matching the given location type. 
The LLM then predicts files based on the prompt. 
The second stage follows the same procedure as the second stage of the pipeline structure–guided LLM.  


\begin{table*}
    \centering
    \small
    \resizebox{1\textwidth}{!}{

\begin{tabular}{ll ccccc ccccc ccccc}
        \toprule
\multirow{2}{*}{Type} & \multirow{2}{*}{LLM} & \multicolumn{5}{c}{File(\%)} & \multicolumn{5}{c}{Class(\%)} & \multicolumn{5}{c}{Function(\%)}\tabularnewline
        \cmidrule(lr){3-7} \cmidrule(lr){8-12} \cmidrule(lr){13-17}

 &  & Acc@1 & Acc@5 & P & R & F1 & Acc@1 & Acc@5 & P & R & F1 & Acc@1 & Acc@5 & P & R & F1\tabularnewline
\midrule 
\midrule
\multirow{2}{*}{ClosedBook} & GPT-4o-mini & 10.4 & 15.1 & 15.8 & 10.0 & 12.3 & 6.8 & 7.9 & 9.0 & 4.4 & 5.9 & 6.1 & 8.6 & 5.8 & 2.3 & 3.3\tabularnewline
 & Claude3.5 & 14.5 & 20.0 & 23.7 & 15.2 & 18.5 & 8.9 & 9.5 & 17.7 & 4.3 & 6.9 & 8.6 & 10.2 & 8.9 & 2.8 & 4.3 \tabularnewline
\midrule
\multirow{2}{*}{ProStructure} & GPT-4o-mini & 12.4& 17.9& 17.5& 12.9& 14.9& 7.9& 9.1& 8.3& 6.9& 7.6& 6.2& 9.1& 4.9& 3.9& 4.3\tabularnewline
 & Claude3.5 & 17.4& 27.0& 27.4& 18.6& 22.1& 9.3& 10.5& 23.9& 8.3& 12.4& 8.6& 11.0& 11.0& 3.9& 5.8\tabularnewline
\midrule 
\multirow{2}{*}{Hint} & GPT-4o-mini & 13.5& 22.3& 16.6& 15.6& 16.1& 6.8& 10.1& 6.3& 7.4& 6.8& 6.5& 9.9& 3.8& 3.6& 3.7\tabularnewline
 & Claude3.5 & 18.6& 32.5& 30.6& 20.2& 24.4& 11.2& 14.1& 17.3& 8.6& 11.5& 9.8& 13.1& 9.0& 5.1& 6.5
 \tabularnewline 
\midrule 
\multirow{2}{*}{Pipeline-ProStructure} & GPT-4o-mini & 10.4& 14.9& 14.5& 9.9& 11.8& 6.6& 7.6& 11.9& 5.2& 7.3& 3.2& 4.8& 6.3& 3.0& 4.1\tabularnewline
 & Claude3.5 & 16.9& 23.3& 27.2& 16.9& 20.9& 10.9& 12.0& 19.3& 8.7& 12.0& 8.1& 9.4& 12.6& 5.9& 8.0\tabularnewline
\midrule 
\multirow{2}{*}{Pipeline-Hint} & GPT-4o-mini & 10.7& 17.3& 16.7& 11.7& 13.8& 6.8& 8.5& 13.4& 5.9& 8.2& 4.2& 6.1& 7.8& 3.4& 4.8\tabularnewline
 & Claude3.5 & 17.7& 24.9& 28.6& 18.5& 22.5& 12.6& 14.5& 26.2& 10.2& 14.7& 8.6& 11.2& 14.5& 6.5& 9.0\tabularnewline

        \bottomrule

\end{tabular}

    }
    \caption{Performance comparison with LLM-based approaches on full \tool}
    \vspace{-.5cm}
    \label{tab:rq2_acc_comparison}
\end{table*}

\subsection{Result} 

\mysec{Approach Comparison.} Table~\ref{tab:rq2_acc_comparison} shows the performance of the LLM-based methods on the full \tool at three localization levels: file, class, and function. 
\textbf{Location-Hint + Claude
3.5} achieves the strongest results across all levels, with only 18.6\% Acc@1, 32.5\% Acc@5, and 24.4\% F1 at the file level, and degraded performance at the class level (14.1\% Acc@5, 11.5\% F1) and function level(13.1\% Acc@5, 6.5\% F1). 
Performance improves from Closed-Book to Project-Structure to Location-Hint. 
Pipeline strategies show mixed results: both weaken file-level accuracy but can improve precision–recall balance at the class and function levels by narrowing predictions to earlier-identified files. 
Overall, project context and file hints clearly enhance localization, while pipelines are not uniformly effective since accurate localization requires integrating information across levels rather than following staged processes. 
The result analysis of each LLM-based approach is provided in Appendix~\ref{rq2_compare_llm}.

\mysec{Model Comparison.} 
Across all approaches and granularities, Claude 3.5 consistently outperforms GPT-4o-mini. 
At the file level, Claude 3.5 improves Acc@5 by 4.9\%–9.8\% and F1 by 5.8\%–9.1\%. 
At the class and function levels, although absolute values are lower, Claude3.5 still yields 1.4\%–6.0\% gains in Acc@5 and 1.0\%–4.2\% gains in F1, demonstrating stronger robustness in fine-grained localization. 

\mysec{Peformance Comparison across Issue Types and Reasons.} Figure~\ref{fig:iss_type_rq2} and Figure~\ref{fig:iss_reason_rq2} shows the performance comparison of different issue types and reasons on the
 full \tool. 
 The performance differences observed on Python code closely mirror those on the full benchmark. 
 The details can seen in Appendix~\ref{rq2_compare_iss_location}.

\mysec{Performance Comparison across Location Scopes.}
Figure~\ref{fig:iss_loc_scope_rq2} presents the performance comparison across different location scopes on the full \tool.  
\textbf{In-Project Files} show the strongest performance, with up to 34\% Acc@5 and 22\% F1 at the file level, and non-negligible results  at finer granularities (14\% Acc@5 and 13\% F1 at the class level and 13\% Acc@5 and 8.5\% F1 at the function level).  
\textbf{User-Authored Files} reach up to 25\% Acc@5 at the file level, while all other metrics remain zero. 
\textbf{Runtime Files} yield limited performance, with the best file-level results around 10\% Acc@5 and F1. 
\textbf{Third-Party Files} perform worst, with all metrics at zero, as external dependencies are rarely referenced in issue descriptions and lack sufficient project-specific context. 
These findings indicate that LLM-based localization is effective on project-maintained code but struggles with external or indirect files.

\mysec{Performance Comparison across Location Types.}  
Figure~\ref{fig:iss_loc_type_rq2} presents the performance comparison across different location types on the full \tool.  
Overall, Code and Configuration Files achieve stronger performance than the other types: for \textbf{Code Files}, ProjectStructure reaches up to 41\% Acc@5 and 30\% F1 at the file level, and Location-Hint achieves 18\%–26\% Acc@5 at class and function levels. 
Notably, Location-Hint performs slightly lower than ProjectStructure at the file level, with a gap of about 2\%. 
This difference is reasonable because Location-Hint distributes its strength more evenly across various file types (e.g., configurations and tests), rather than being specialized for source code as ProjectStructure is. 
For \textbf{Configuration Files}, Location-Hint achieves the best results with 30\% Acc@5 and 26\% F1 at the file level.  
\textbf{Test Files} show moderate performance, with Location-Hint achieving 29\% Acc@5 and 22\% F1 at the file level, but class- and function-level F1 often falls below 5\% due to very low recall. 
\textbf{Documentation Files} yield limited performance, with the best results around 22\% Acc@5 and 20\% F1.  
\textbf{Asset Files} show similarly limited result, with 20\% Acc@5 and 21\% F1.  
These results indicate that LLM-based localization is more effective for code-related files, whereas performance drops for non-code files. 
\vspace{-1.8mm}
\section{Related Work}\label{relatedwork}


\noindent \textbf{Project Benchmarks for Evaluating LLMs.}  
With the rise of large language models (LLMs), several benchmarks have emerged to evaluate their performance~\cite{jimenez2023swe,chen-etal-2025-locagent,zhuo2024bigcodebench,deng2025nocode,niu2023crosscodebench,chen2021evaluating,ouyang2024benchmarking,gao2023benchmarking,jiang2024collubenchbenchmarkpredictinglanguage,jain2024r2e,mundler2024swt,xie2024osworld}.  
SWE-Bench~\cite{jimenez2023swe}, collects 2,294 issues with pull requests from 12 Python projects, mainly focusing on bug fixing.  
SWE-Bench Lite provides a subset of 300 issues to encourage adoption, which is widely used as a standard benchmark for research. 
Recently, LocBench~\cite{chen-etal-2025-locagent} expands coverage to 560 issues with more diverse issue types. 
These benchmarks primarily focus on Python code and rely heavily on pull requests as the sole resolution signal.  
In reality, software project issues can be addressed through pull requests, commits, or comments.  
To better reflect these real-world scenarios, our \tool collects 1,100 issues accompanied by pull requests, commits, or comments, offering broader and more balanced coverage across issue types, causes, location scopes, and types. 

\noindent \textbf{Issue Localization.}  
Issue localization refers to the set of code or non-code artifacts (e.g., source files, test files, configuration files, documentation) that are relevant for resolving software issues.  
Traditional retrieval-based methods~\cite{robertson1994micheline,wang2022text,suresh2024cornstack}, such as BM25, rely on lexical matching to return a ranked list of potentially relevant files.  
Recently, LLM-based retrieval methods have been proposed to improve localization performance~\cite{xia2024agentless,chen-etal-2025-locagent,reddy2025swerank,wang2024openhands,yang2024swe,jiang2025cosil,zhang2024autocoderover,tao2024magis,xie2025swe, ma2025sorft}.  
For example, Agentless~\cite{xia2024agentless} adopts a three-phase approach to identify relevant code locations.  
LocAgent~\cite{chen-etal-2025-locagent} further introduces a heterogeneous graph representation to enhance code localization. 
Currently, LocAgent achieves over 70\% top-1 accuracy on file-, class-, and function-level localization on SWE-Bench Lite.  
However, when evaluated on \tool, LocAgent achieves less than 40\% Acc@5. 
This highlights the increased difficulty and broader diversity of our benchmark, as well as the limitations of existing methods in generalizing to more realistic and complex issue scenarios.


\vspace{-1.8mm}
\section{Conclusion}\label{conclusion}
In this paper, we present \tool, a comprehensive project location benchmark for issue resolution. 
Compared with existing benchmarks, \tool covers more diverse issue types, root causes, location scopes, and file types, providing a richer basis for evaluation. 
Our experiments with state-of-the-art and LLM-based approaches reveal that even the best methods achieve below 40\% Acc@5 at the file level, highlighting substantial challenges and the need for further advances in realistic issue resolution. 


\bibliography{iclr2026_conference}
\bibliographystyle{iclr2026_conference}

\appendix
\section{Appendix}
\subsection{Issue Filtering Process} \label{issue_filter_process}
\textbf{Issue filtering:} Reliable location information is critical for our study, but many issues lack such information due to being open, unresolved, or having ambiguous descriptions. 
To obtain reliable location data, we apply three filtering conditions: selecting closed issues, retaining those with confirmed resolutions, and keeping only issues with clear location information. 
After filtering,  there are 1,100 issues from 46 projects. 
Among them, 716 have pull requests, 63 have commits, and 321 have resolution-related comments. 
The details are as follows:  

\begin{itemize}[left=2pt, labelsep=1mm,leftmargin=*,labelindent=4pt]

    \item \textit{Select closed issues:} We exclusively consider closed issues, as open issues typically mean ongoing discussions or incomplete resolutions. 
    \item \textit{Determine resolution status:} For closed issues, we further identify whether they are resolved.
We observe three common ways to determine resolution status: 
 \begin{itemize}[labelsep=1mm,leftmargin=*,labelindent=4pt]
\item  The issue has a corresponding merged pull request (PR). 

\item  The issue is referenced or closed by a commit that has been integrated into the main branch. 

\item  The issue comments include confirmation, such as ``thank you, it answered my question,'' indicating the issue has been resolved. 
 \end{itemize}   
     \item \textit{Check location clarity:} We perform this check only for issues without linked pull requests or commits, as those contain location information. 
For such issues, we analyze whether the comments explicitly state the location relevant to resolving the issue (e.g., ``got it working, just changed \dots{} in \texttt{ui.py}'').
\end{itemize}

\subsection{Empirical Analysis of Issues and Locations} 

\subsubsection{Analysis Method}\label{analysis_method_iss_location}
In Section~\ref{method}, we collect 1,100 issues associated with pull requests, commits, or resolution-related comments. 
 To define taxonomy for issues or locations, we employ the card sorting approach with two iterations~\cite{readability_idioms_zhang,zheng2025towards_card_sorting, bi2022accessibility,liu2024refining}. 
In the first iteration, we randomly sample issues for each type with a confidence level of 95\% and an error
margin 5\%. 
Then, two authors first independently analyze the issue report and then they annotate the issue with a short description. 
They discuss and resolve all disagreements if
their descriptions do not have the same meaning. 
The Cohen's kappa agreement between two labels all above 0.7 (substantial agreement). 
Next, they work together to group all annotations into a  category with the corresponding definition. 

In the second iteration, two authors independently annotate the remaining issues with the predefined categories. 
If the remaining issues do not fit existing categories, they are annotated with new short descriptions. 
It is found that no new categories are needed. 
The Cohen's kappa agreement for issue and location category between two labels all above 0.75. 
Then, two authors
discuss the disagreements to reach an agreement. 
Figure~\ref{fig:comparison_iss_num} shows examples of issues with different types, root causes, location scopes, and location types. 

\subsubsection{Result} \label{analysis_result_iss_location}
\noindent \textbf{Issue Type} captures the kinds of problem developers raise in issue reports, such as errors during code execution. 
It reflects what challenges developers encounter or what support they seek during project use and maintenance. 
We categorize issues into four types. 

\begin{enumerate}[label={\bfseries(\arabic*)}, left=0pt, labelsep=1mm, leftmargin=*, labelindent=0pt]
\item \textbf{Execution Failure}: Issues where the project fails to run properly, such as installation errors, startup crashes, or runtime errors. 
This category accounts for 39.9\% (439 issues). 
For example, in panel (a) of Figure~\ref{fig:comparison_iss_num}, a developer reports  an AttributeError: module `torch.cuda' has no attribute `comm' when executing code with the \texttt{pytorch} project.

\item \textbf{Unexpected Result}: Issues where the project runs successfully, but its behavior or output deviates from expectations, such as producing incorrect results. 
It accounts for 23.7\% (261 issues). 
For example, in panel (c) of Figure~\ref{fig:comparison_iss_num}, a developer attempts to run \texttt{Fooocus} to load the model, but the program simply exits without loading it or showing any errors. 

\item \textbf{Enhancement Request}: Issues where users request new functionality, improvements to existing features, or extensions to the project's capabilities. 
It accounts for 25.1\% (276 issues). 
For example, in panel (b) of Figure~\ref{fig:comparison_iss_num}, a developer requests remove unnecessary configurations, such as tdd. 

\item \textbf{Usage Question}: Issues where users ask about project usage, such as how to accomplish specific tasks, locating files, or understanding code structure. 
It accounts for 11.3\% (124 issues). 
For example, in panel (d) of Figure~\ref{fig:comparison_iss_num}, a developer asks how to use \texttt{Sponsorblock} as part of Python script. 


\end{enumerate}

\noindent \textbf{Root Causes of Issues} 
capture the underlying reasons behind issue reports, such as code bugs. 
We categorize issue causes into three types: 

\begin{enumerate}[label={\bfseries(\arabic*)}, left=0pt, labelsep=1mm, leftmargin=*, labelindent=0pt]

\item \textbf{Implementation Bug}: Problems arising due to errors in the project's implementation, where the functionality does not perform as expected, such as logic errors and incorrect outputs.  
It accounts for 34.5\% (380 issues). 
For example, in panel (a) of Figure~\ref{fig:comparison_iss_num}, a developer fixes an AttributeError raised by calling a non-existent API \texttt{torch.cuda.comm.broadcast}. 

\item \textbf{Design Deficiency}: Situations where a project is incomplete or poorly implemented, such as missing functionality and need for refactoring. 
It accounts for 36.3\% (399 issues). 
For example, in panel (b) of Figure~\ref{fig:comparison_iss_num}, a developer removes unnecessary configurations to improve design clarity and maintainability. 

\item \textbf{User-Induced Problem}: Problems arising from incorrect usage, misunderstandings, or limited knowledge of the project on the user's side. 
It accounts for 29.2\% (321 issues). 
For example, in panel (c) of Figure~\ref{fig:comparison_iss_num}, developers fail to load a model in \texttt{Fooocus} until they correct their own configuration settings. 

\end{enumerate}

\noindent \textbf{Location Scope for Issue Resolution} refers to the range of files involved in issue resolution, which can also include locations beyond the project repository. 
To use a software project, developers depends not only on files within the project itself, but also on external dependencies, runtime-generated files, and user-authored code. 
We classify location scopes into four categories:  
 
\begin{enumerate}[label={\bfseries(\arabic*)}, left=0pt, labelsep=1mm, leftmargin=*, labelindent=0pt]

\item \textbf{In-Project File}: Files that are physically present within the project repository.  
Typical examples include source code files (\texttt{main.py}), configuration files (\texttt{settings.yaml}), or documentation (\texttt{README.md}).  
It accounts for 94.1\% (1035 issues). 
For example, in panel (b) of Figure~\ref{fig:comparison_iss_num}, an issue from the \texttt{AntonOsika/gpt-engineer} project was resolved by modifying three files within the repository, which shows a case of in-project file localization. 

\item \textbf{Runtime File}: Files that are outside the project but are generated during execution or required at runtime. 
It accounts for 2.2\% (24 issues). 
For example, a generated file created after running a source script, or an environment-specific configuration file such as \texttt{.env}. 
In panel (c) of Figure~\ref{fig:comparison_iss_num}, an issue from the \texttt{lllyasviel/Fooocus} project falls into this category: its resolution required modifying two files, \texttt{config.txt} and \texttt{config\_modification\_tutorial.txt}, which are not part of the repository but are created after execution. 


\item \textbf{Third-Party File}: Files from external projects, such as third-party libraries that the current project depends on. 
It accounts for 2.1\% (23 issues). 
For example, in panel (a) of Figure~\ref{fig:comparison_iss_num}, an issue reported in the \texttt{pytorch/pytorch} project is ultimately traced to an implementation bug in the external dependency \texttt{facebookresearch/InterHand2.6M}, rather than in PyTorch itself. 

\item \textbf{User-Authored File}: Files written by users outside the original project, typically referenced in issue descriptions when reporting problems. 
It accounts for 3.0\% (33 issues). 
For example, in issue 78759 of the \texttt{ansible/ansible} project, a user reports an error when passing a variable to a loop and provides the corresponding configuration. 
Another user suggests to replace \texttt{loop:\{\{sysctl.values\}\}} with \texttt{loop:\{\{sysctl[`values']\}\}}, which resolves the issue. 
In this case, the location of the issue resolution is the user's own configuration file, specifically the line \texttt{loop:\{\{sysctl.values\}\}}. 


\end{enumerate}

\noindent \textbf{Location Type for Issue Resolution} refers to the file types affected during issue resolution. 
A software project is not solely code; it also depends on configurations, documentation, and other artifacts that together to function as a complete project. 
We categorize issue locations into five types based on their functional roles.  




\begin{enumerate}[label=\bfseries(\arabic*), left=0pt, labelsep=1mm, leftmargin=*, labelindent=0pt] 

\item \textbf{Code}: Source files that implement the core logic and functionality of the project, typically written in programming languages such as \texttt{.py} or \texttt{.cpp}. 
It accounts for 80.8\% (889 issues). 
For example, in panel (a) of Figure~\ref{fig:comparison_iss_num}, the resolution was located in \texttt{common/nets/module.py}, which is a Python source file. 

\item \textbf{Test}: Test files that verify the correctness of the project through test cases. 
This category includes unit, integration, and system test files, e.g., \texttt{test\_*.py} and \texttt{*.spec.js}. 
It accounts for 23.5\% (258 issues). 
For example, in panel (b) of Figure~\ref{fig:comparison_iss_num}, the issue resolution location involves the file \texttt{tests/test\_collect.py}, which is a Python test file. 

\item \textbf{Configuration}: Configuration files used for build processes, dependency management, or runtime settings, such as \texttt{requirements.txt} and \texttt{pom.xml}. 
It accounts for 15.2\% (167 issues). 
For example, in panel (c) of Figure~\ref{fig:comparison_iss_num}, the issue resolution location contains the file \texttt{presets/default.json}, which is a configuration file. 


\item \textbf{Document}: Documentation files intended to describe and support the usage and maintenance of the project, typically found in files such as \texttt{README} or within the \texttt{docs} directory. 
It accounts for 23.5\% (259 issues). 
For example, in panel (d) of Figure~\ref{fig:comparison_iss_num}, the issue resolution location involves \texttt{README.md}, which provides guidance for developers on using the project. 


\item \textbf{Asset}: Non-code resources that support system operation or enhance functionality, such as data files (e.g., \texttt{.csv}, \texttt{.json}), static resources (e.g., images, fonts), or auxiliary scripts. 
It accounts for 4.4\% (48 issues). 
For example, in panel (b) of Figure~\ref{fig:comparison_iss_num}, the issue resolution location involves \texttt{gpt\_engineer/preprompts/spec}. 
As this file contains only prompt text consumed as input, rather than executable code, it is classified as an asset.
\end{enumerate}

\subsection{State-of-Art Approaches Comparison}\label{rq1_approach_comparison}
From Table~\ref{tab:rq1_acc_comparison}, the four approaches show different performance: LocAgent and OpenHands yield the best and comparable results, followed by Agentless and then BM25. 
The details are as follows: 
\begin{itemize}[left=2pt, labelsep=1mm,leftmargin=*,labelindent=4pt]

\item \textbf{BM25:} Yields the weakest results across all levels.  
Acc@5 remains below 10\%, and F1 stays under 14\% for all levels. 
This confirms that lexical retrieval alone struggles with reasoning and semantic understanding. 


\item \textbf{Agentless:} Provides moderate improvements over BM25, with the best file-level results of 23.9\% Acc@5 and 17.0\% F1. 
However, class-level performance remains weak, with gains over BM25 limited to less than 2\% Acc@5 and in some cases even lower (e.g., 4.5\% Acc@5), indicating instability across granularities.  


\item \textbf{LocAgent:} Offers the best overall performance, especially when combined with Claude 3.5.  
It reaches 35.2\% Acc@5 and 38.1\% F1 at the file level, while maintaining around 20\% Acc@5 and F1 at the class and function levels.  
The results suggest that incorporating graph-based state representations helps enhance reasoning and improves localization across granularities.  

\item \textbf{OpenHands:} Reaches performance close to LocAgent with Claude 3.5. 
At the file level, the differences in Acc@5 and F1 remain within 4\%. 
At the class and function levels, its Acc@5 and F1 are even slightly higher (within 3\%). 
Notably, OpenHands is less sensitive to model choice, as the gap between Claude 3.5 and GPT-4o-mini is smaller than that observed for LocAgent. 
\end{itemize}

\subsection{Performance comparison of LLM-based approaches on Full \tool} \label{rq2_compare_llm} 
Table~\ref{tab:rq2_acc_comparison} shows the performance of the LLM-based methods on the full \tool at three localization levels: file, class, and function. 
\textbf{Location-Hint + Claude
3.5} achieves the strongest results across all levels, with only 18.6\% Acc@1, 32.5\% Acc@5, and 24.4\% F1 at the file level, and degraded performance at the class level (14.1\% Acc@5, 11.5\% F1) and function level(13.1\% Acc@5, 6.5\% F1). 
Performance improves from Closed-Book to Project-Structure to Location-Hint. 
Pipeline strategies show mixed results: both weaken file-level accuracy but can improve precision–recall balance at the class and function levels by narrowing predictions to earlier-identified files. 
Overall, project context and file hints clearly enhance localization, while pipelines are not uniformly effective since accurate localization requires integrating information across levels rather than following staged processes. 
The details are as follows: 

\begin{itemize}[left=2pt, labelsep=1mm,leftmargin=*,labelindent=4pt]

\item \textbf{Closed-Book:} Performs the weakest. At the file level, Claude 3.5 reaches only 20.0\% Acc@5 and 18.5\% F1.  
At class and function levels, performance drops further, with Acc@5 below 10.2\% and F1 below 7\%, showing the difficulty of relying solely on prior knowledge without project context.  

\item \textbf{Project-Structure:} Adds clear gains at the file level (27.0\% Acc@5, 22.1\% F1), improving by +7.0\% Acc@5 and +3.6\% F1 over Closed-Book.  
At the class level, improvements are modest (10.5\% Acc@5, 12.4\% F1), while function-level benefits are minimal (11.0\% Acc@5, 5.8\% F1).  


\item \textbf{Location-Hint:} Achieves the best overall results. 
At the file level, Claude 3.5 reaches 32.5\% Acc@5 and 24.4\% F1, outperforming Project-Structure by +5.5\% and +2.3\%.  
At the class level it achieves 14.1\% Acc@5 and 11.5\% F1 (+3.6\% and –2.3\% vs. Project-Structure), and at the function level 13.1\% and 6.5\% (+1.9\%, +0.7\%). 
Notably, at finer granularities trade-offs appear: Location-Hint is lower in precision but higher in recall than Project-Structure, leading to slightly reduced F1 at the class level. 
These results show that file hints are most effective at the file level, while their benefit diminishes at finer granularities, sometimes trading precision for recall.  

\item \textbf{Pipeline-ProStructure:} Its effectiveness does not surpass that of Project-Structure. 
At the file level, Claude 3.5 reaches only 23.3\% Acc@5 and 20.9\% F1, lower than Project-Structure (27.0\% Acc@5, 22.1\% F1).  
At the class and function levels, Acc@5 also degrades, but a precision–recall trade-off emerges, making the F1 comparable to, and in some cases slightly higher than, Project-Structure.  
This suggests that pipelines dilute file-level benefits and do not consistently improve finer-grained localization.  


\item \textbf{Pipeline-Hint:} More stable than Pipeline-ProStructure, though weaker than direct Hint at the file level (24.9\% Acc@5/22.5\% F1 vs. 32.5\%Acc@5/24.4\% F1). 
At the class level (14.5\% Acc@5/14.7\% F1) and function level (11.2\% Acc@5/9.0\% F1), it surpasses direct Hint (14.1\% Acc@5/11.5\% F1, 13.1\% Acc@5/6.5\% F1). 
This suggests that, given more reliable file-level predictions, pipelining can be effective at finer granularities.


\end{itemize}

\subsection{Figures of Performance comparison of different issue types, reasons, location scopes and types on Full \tool} \label{rq2_compare_iss_location}

\begin{figure*}[!h]
  \centering
    \setlength{\abovecaptionskip}{1mm}
    \includegraphics[width=5.5in]{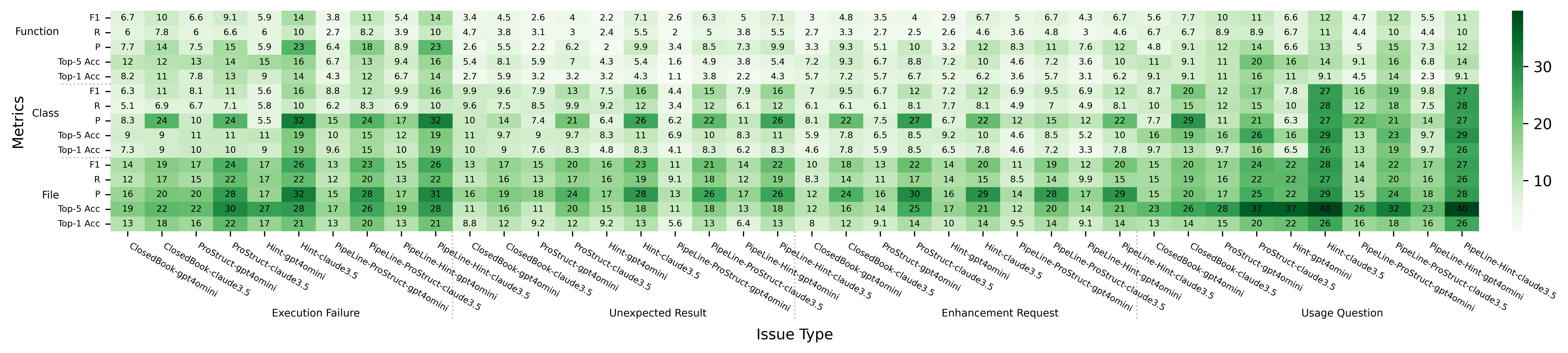}%
    \caption{Performance comparison of different issue types on full \tool. }  
    \label{fig:iss_type_rq2} 
    \vspace{-0.4cm}
\end{figure*}

\mysec{Peformance Comparison across Issue Types.}
Figure~\ref{fig:iss_type_rq2} presents the performance comparison across different issue types on the full \tool. 
The performance differences observed on Python code closely mirror those on the full benchmark. 
\textbf{Execution Failures} achieve moderate results, with Location-Hint reaching 28\% Acc@5 and 26\% F1 at the file level. 
Their performance is relatively balanced across location levels (file, class, function) and metrics (Acc@1, Acc@5, precision, recall, F1), typically yielding mid-range scores: stronger than Enhancement Requests and Unexpected Results but weaker than Usage Questions, indicating a moderate level of localization difficulty. 
\textbf{Unexpected Results and Enhancement Requests} yield the weakest performance, with Acc@5 capped at around 20\% at the file level. 
At the class level, Acc@5 and F1 are around 10\% and 15\%, while at the function level both fall below 10\%, reflecting the challenge of pinpointing subtle or underspecified behavioral issues.  
\textbf{Usage Questions} achieve the strongest overall performance, with Location-Hint reaching 40\% Acc@5 and 28\% F1 at the file level. 
At the class level, Acc@5 and F1 are 29\% and 27\%, and at the function level both remain around 10\%. 
Their relatively higher performance likely due to clearer textual cues in issue descriptions that make localization easier.

\begin{figure*}[!h]
  \centering
    \setlength{\abovecaptionskip}{1mm}
    \includegraphics[width=5.5in]{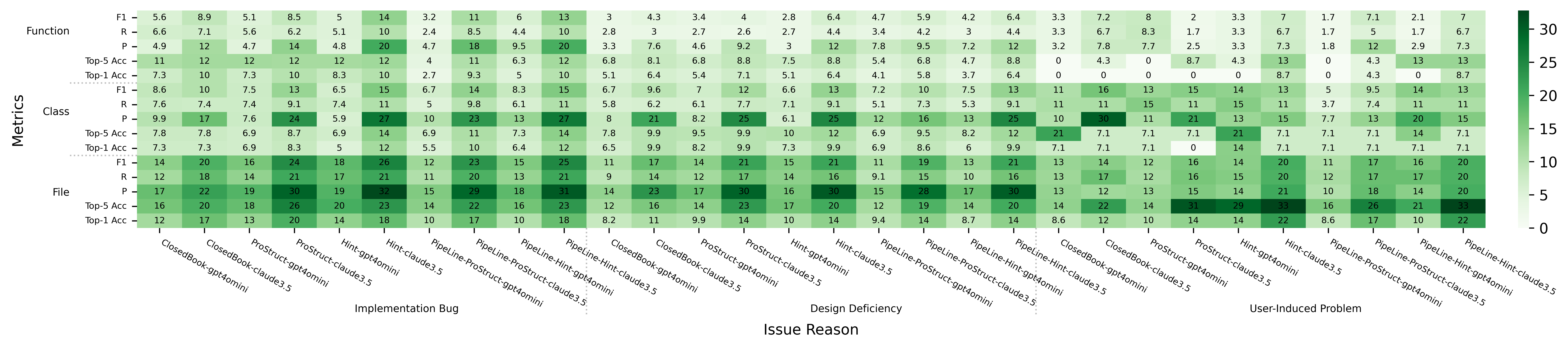}%
    \caption{Performance comparison of different issue reasons on full \tool. }  
    \label{fig:iss_reason_rq2} 
    \vspace{-0.4cm}
\end{figure*}

\mysec{Peformance Comparison across Issue Reasons.} Figure~\ref{fig:iss_reason_rq2} shows the performance comparison of different issue reasons on the
 full \tool. 
 \textbf{Implementation Bug} achieve the most stable performance, with Location-Hint reaching 23\% Acc@5 and 26\% F1 at the file level, and 12–15\% at finer levels. 
Although bug fixing is generally considered highly challenging, the relatively balanced performance across levels, together with the long-standing emphasis of researchers on bug-fix scenarios, explains why this issue reason performs more consistently than others. 
\textbf{Design Deficiencies} perform weakest, with best file-level results of 20\% Acc@5 and 21\% F1, but Acc@5 and F1 at the class and function levels drop to 5.4\%–16\%. 
Notably, precision is obviously higher than recall across all three levels. 
Across all levels, precision is higher than recall, as design-related issues are abstract and cross-cutting: they are harder to locate, but once identified, predictions tend to be precise. 
\textbf{User-Induced Problems} show intermediate performance, with 33\% Acc@5 and 20\% F1 at the file level, but only 7\%–13\% at the class and function levels. 
This suggests that user mistakes often provide clear textual cues that aid file-level localization. 
However, such cues are less informative for finer-grained levels, which results in weaker class- and function-level performance. 

%

\begin{figure*}[!h]
  \centering
    \setlength{\abovecaptionskip}{1mm}
    \includegraphics[width=5.5in]{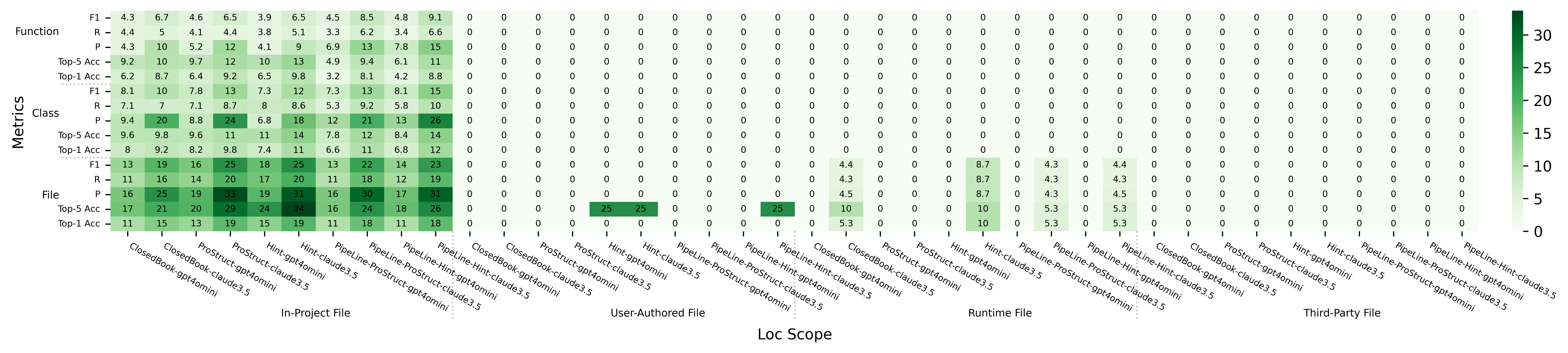}%
    \caption{Performance comparison of different location scopes on full \tool. }  
    \label{fig:iss_loc_scope_rq2} 
    \vspace{-0.4cm}
\end{figure*}

\begin{figure*}[!h]
  \centering
    \setlength{\abovecaptionskip}{1mm}
    \includegraphics[width=5.5in]{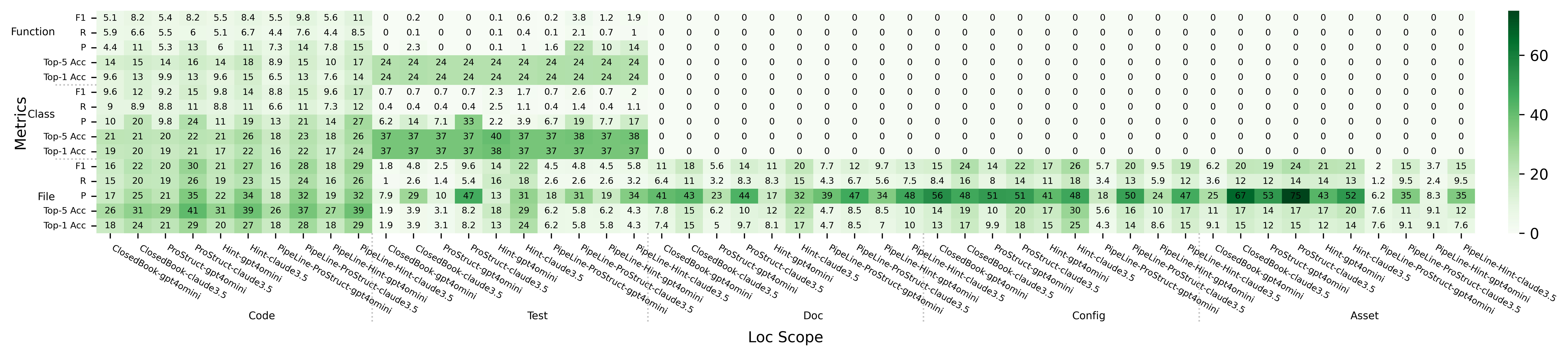}%
    \caption{Performance comparison of different location types on full \tool. }  
    \label{fig:iss_loc_type_rq2} 
    \vspace{-0.9cm}
\end{figure*}

\newpage
\subsection{Prompt Templates of LLM-based Approaches}\label{prompt_template} 
\begin{figure*}[h]
  \centering
    \setlength{\abovecaptionskip}{2mm}
    \includegraphics[width=5.5in]{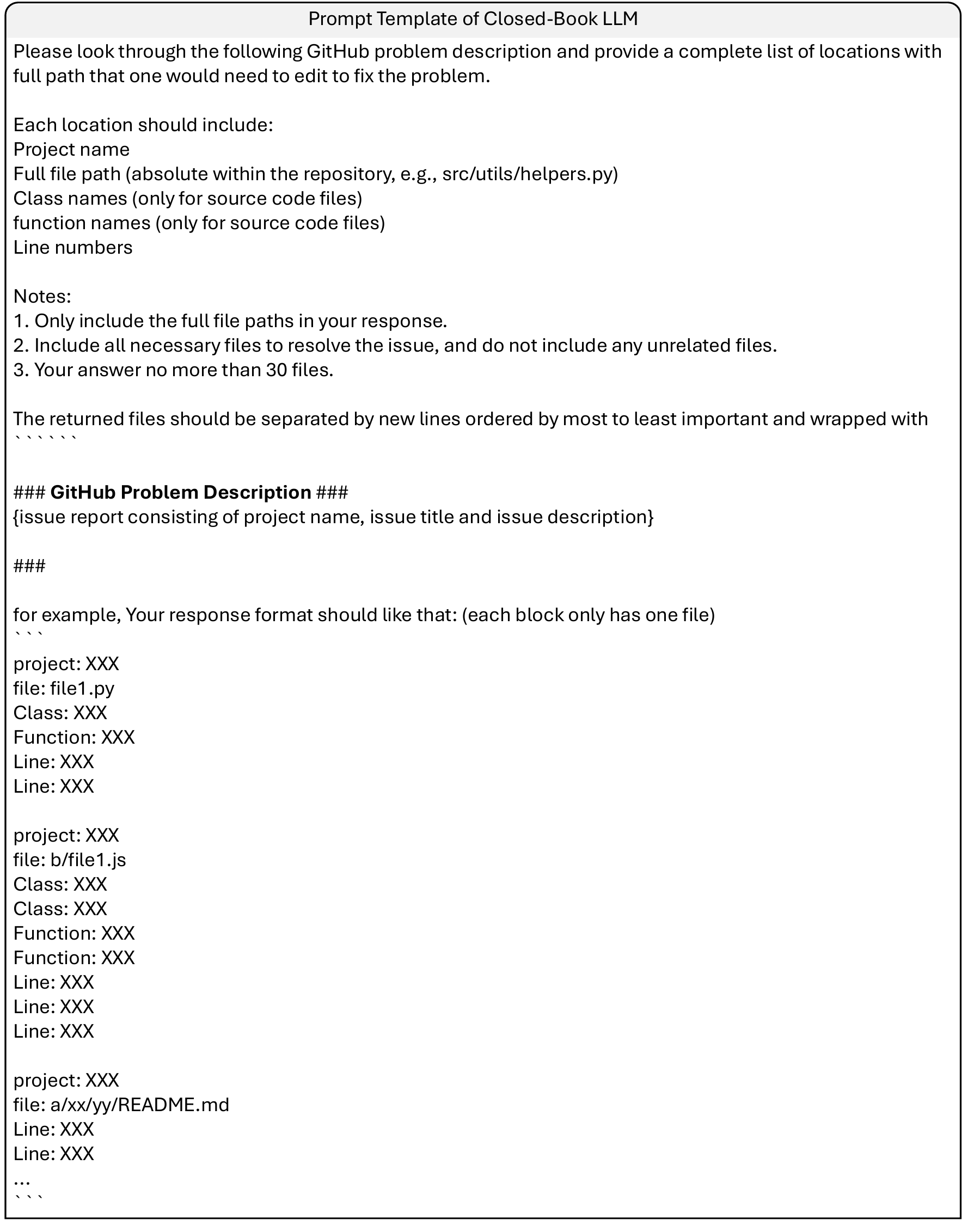}%
    \caption{Prompt template of closed-book LLM in project localization  for issue resolution. }  
    \label{fig:prompt_template_closed-book} 
    \vspace{-0.4cm}
\end{figure*}

\begin{figure*}
  \centering
    \setlength{\abovecaptionskip}{2mm}
    \includegraphics[width=5.5in]{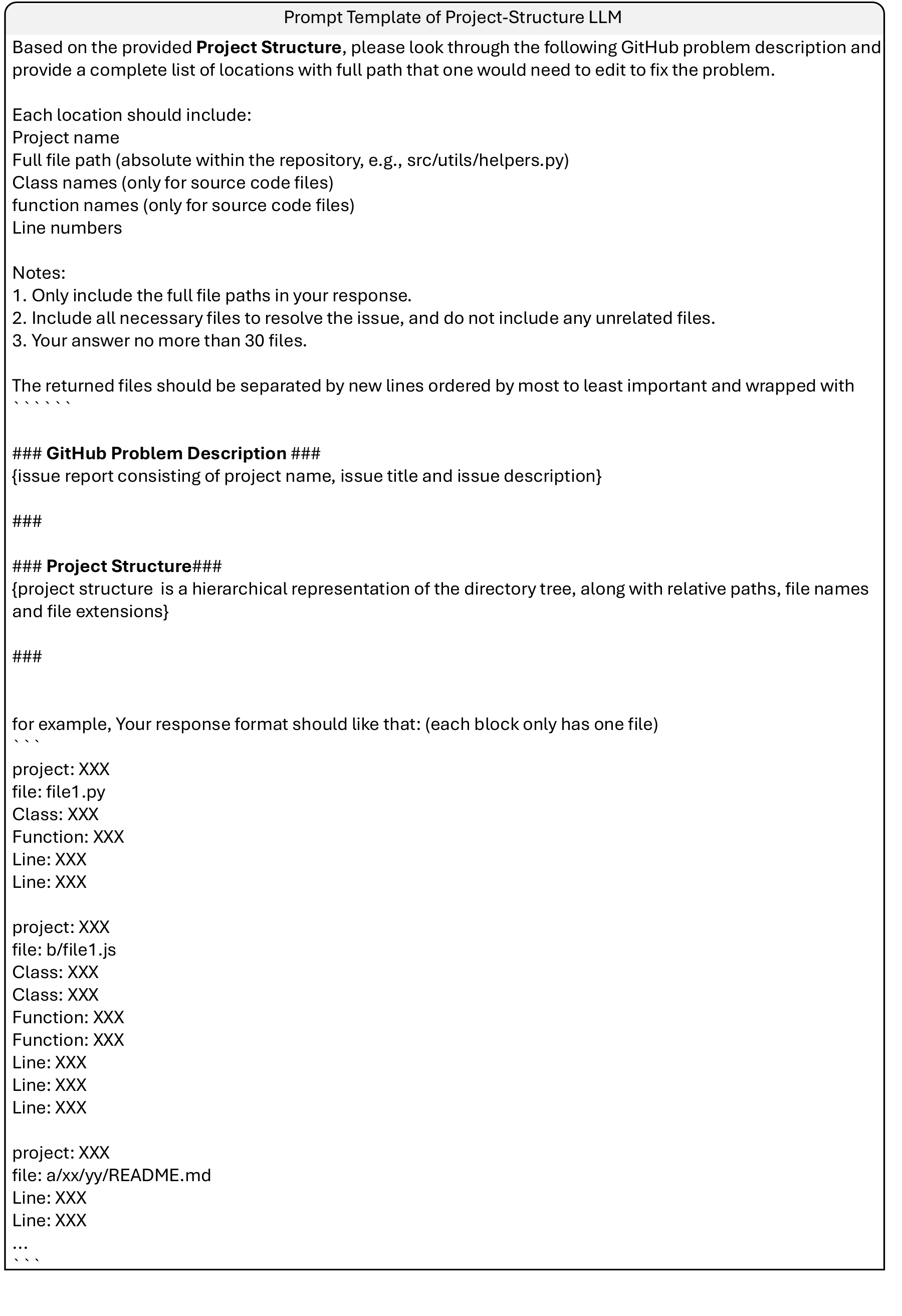}%
    \caption{Prompt template of Project-Structure LLM in project localization  for issue resolution. }  
    \label{fig:prompt_template_prostructure} 
    \vspace{-0.4cm}
\end{figure*}

\begin{figure*}
  \centering
    \setlength{\abovecaptionskip}{-2mm}
    \includegraphics[width=5.1in]{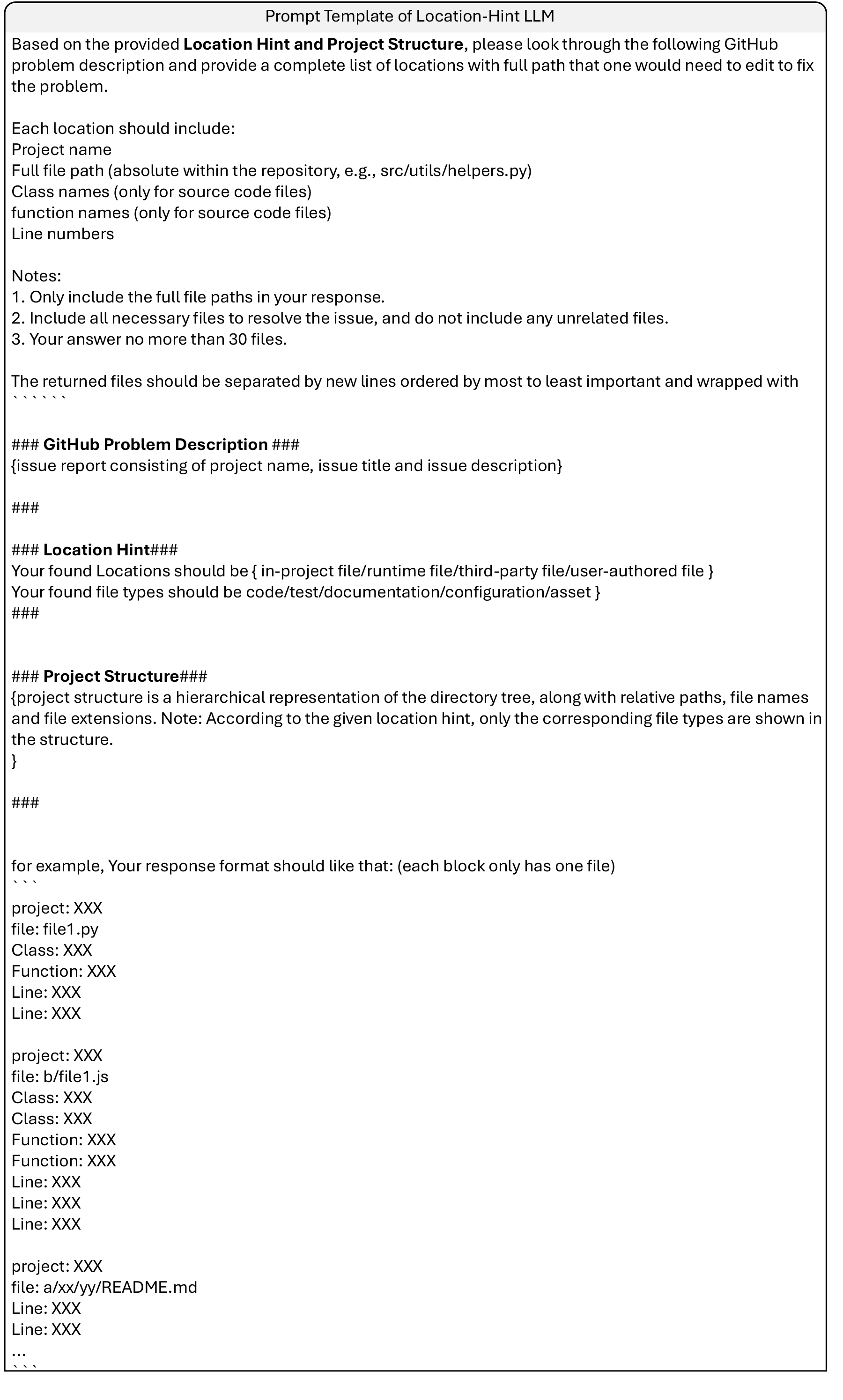}%
    \caption{Prompt template of Location-Hint LLM in project localization for issue resolution. }  
    \label{fig:prompt_template_lochint} 
    \vspace{-0.4cm}
\end{figure*}

\begin{figure*}
  \centering
    \setlength{\abovecaptionskip}{-2mm}
    \includegraphics[width=5.1in]{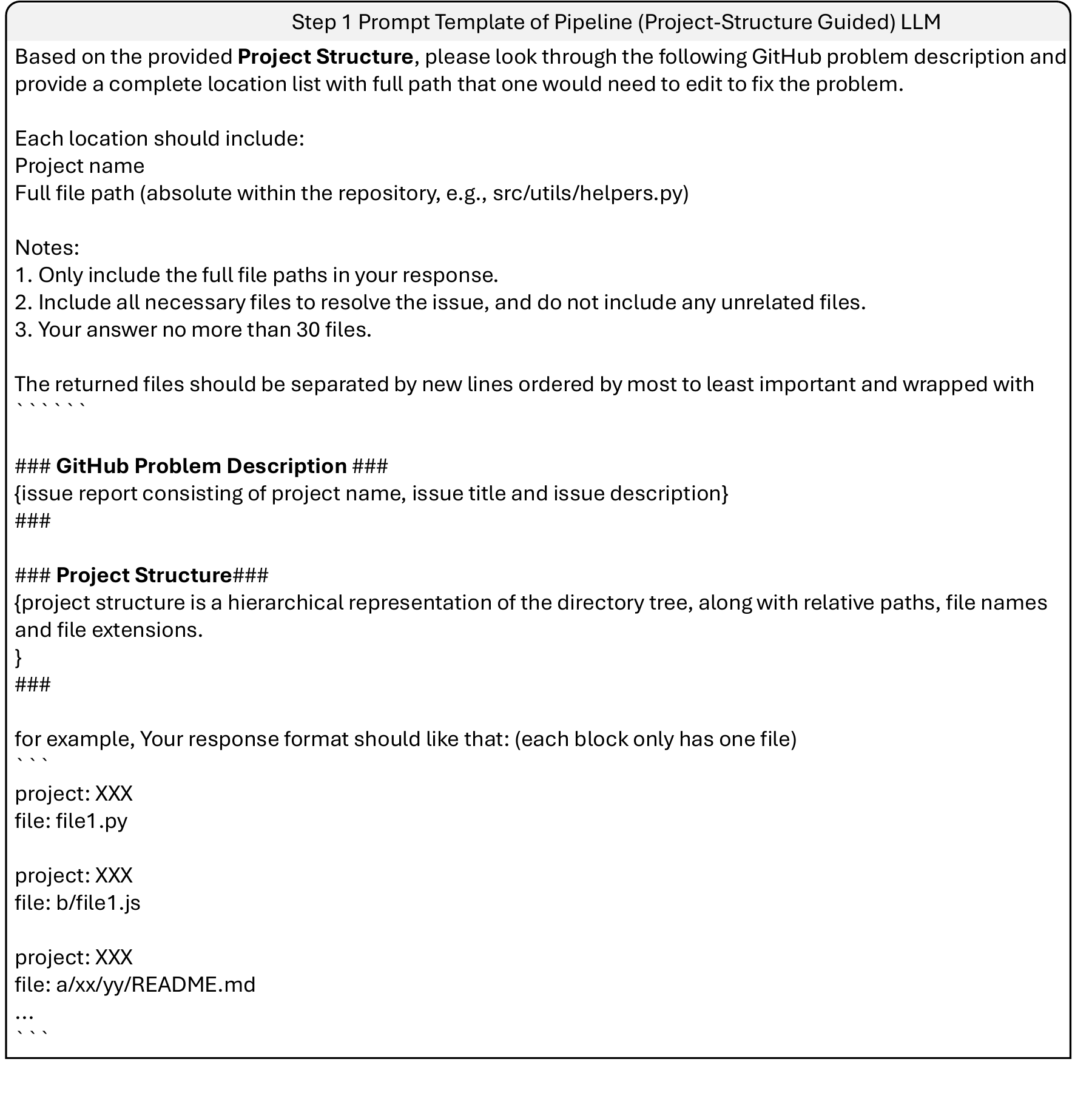}%
    \caption{Step 1 Prompt template of PipeLine (Project-Structure Guided) LLM in project localization for issue resolution. }  
    \label{fig:prompt_template_pipeline1_step1} 
    \vspace{-0.4cm}
\end{figure*}

\begin{figure*}
  \centering
    \setlength{\abovecaptionskip}{-2mm}
    \includegraphics[width=5.1in]{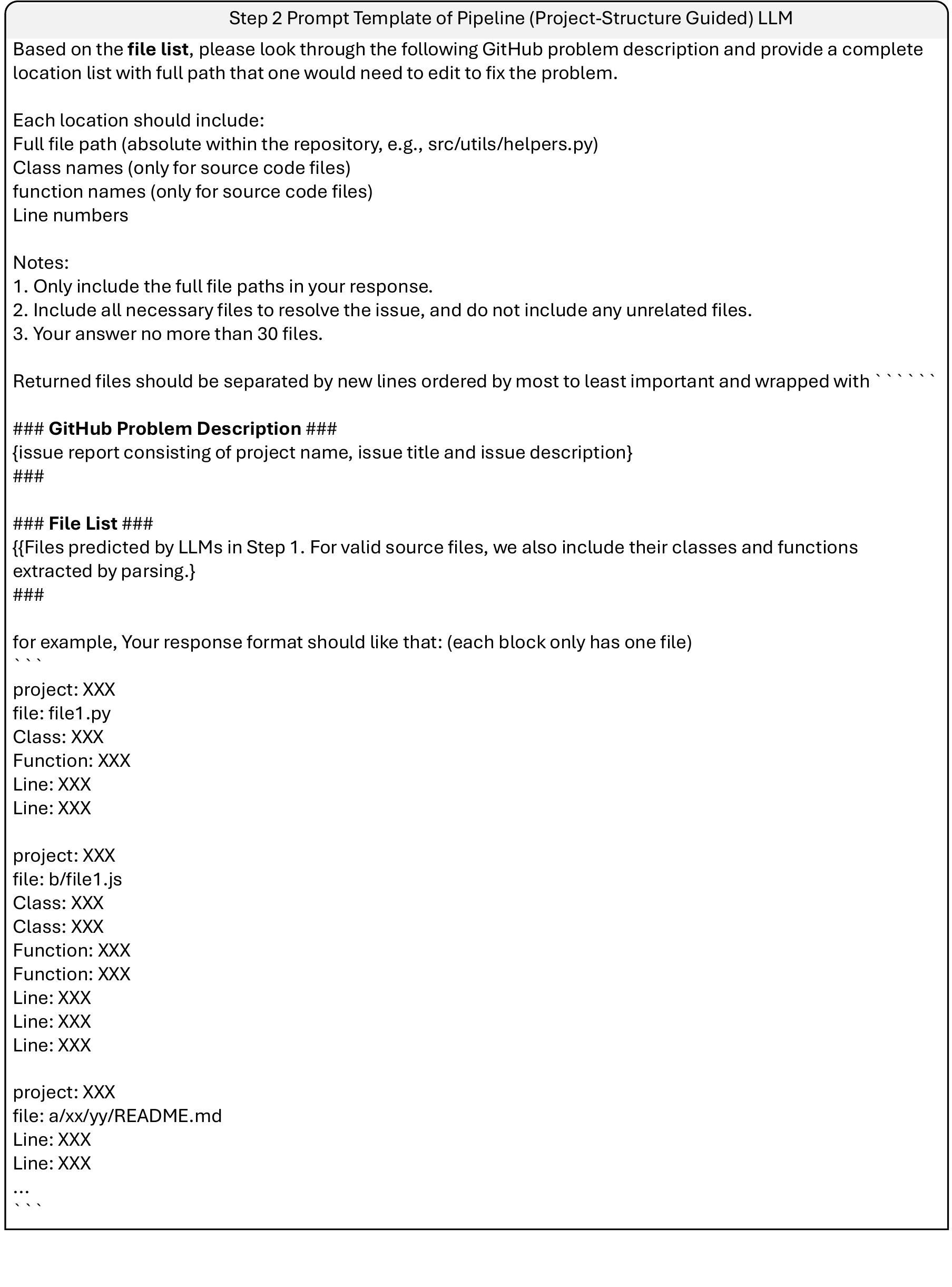}%
    \caption{Step 2 Prompt template of PipeLine (Project-Structure Guided) LLM in project localization for issue resolution. }  
    \label{fig:prompt_template_pipeline1_step2} 
    \vspace{-0.4cm}
\end{figure*}

\begin{figure*}
  \centering
    \setlength{\abovecaptionskip}{-2mm}
    \includegraphics[width=5.1in]{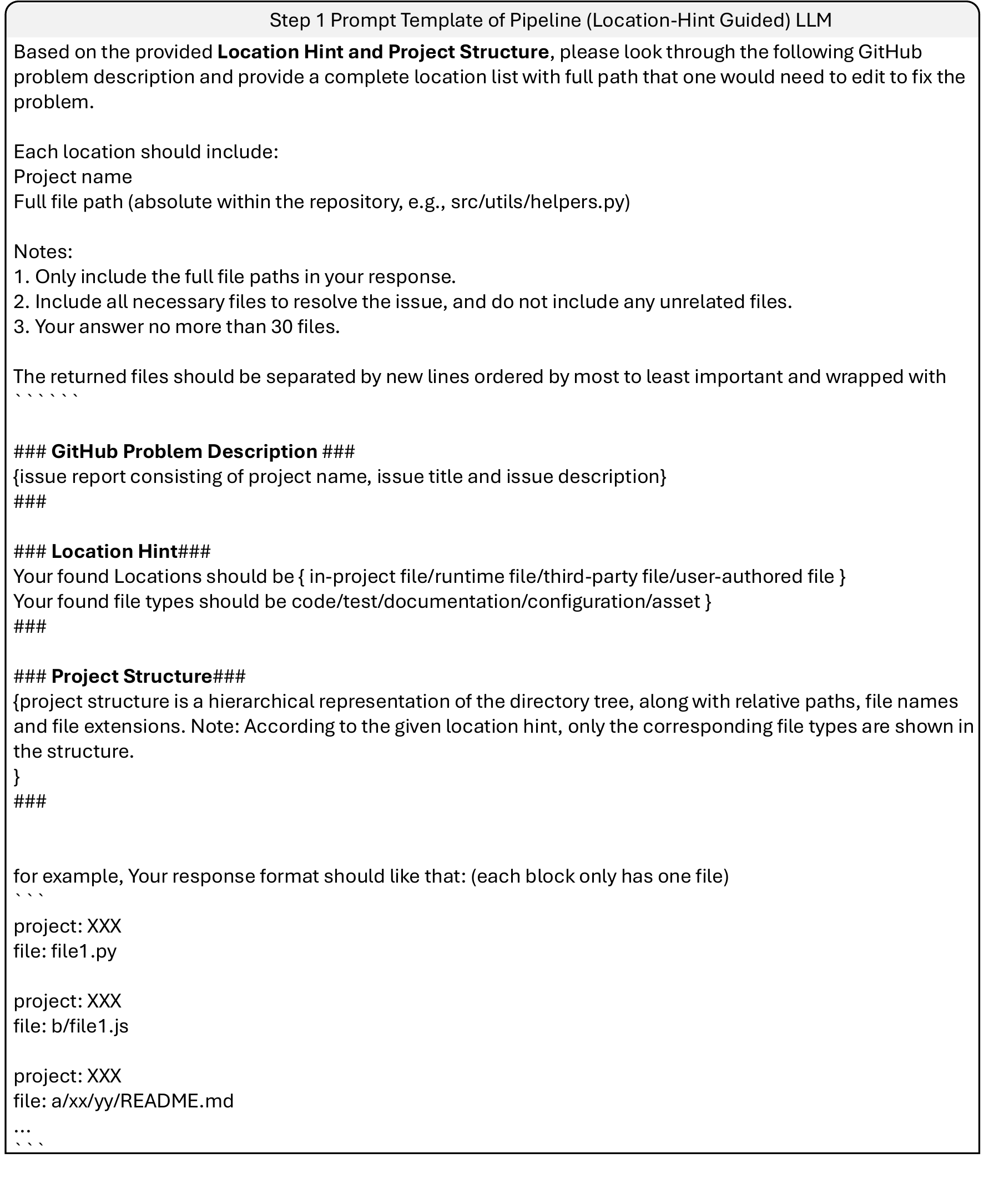}%
    \caption{Step 1 Prompt template of PipeLine (Location-Hint Guided) LLM in project localization for issue resolution. }  
    \label{fig:prompt_template_pipeline2_step1} 
    \vspace{-0.4cm}
\end{figure*}

\begin{figure*}
  \centering
    \setlength{\abovecaptionskip}{-2mm}
    \includegraphics[width=5.1in]{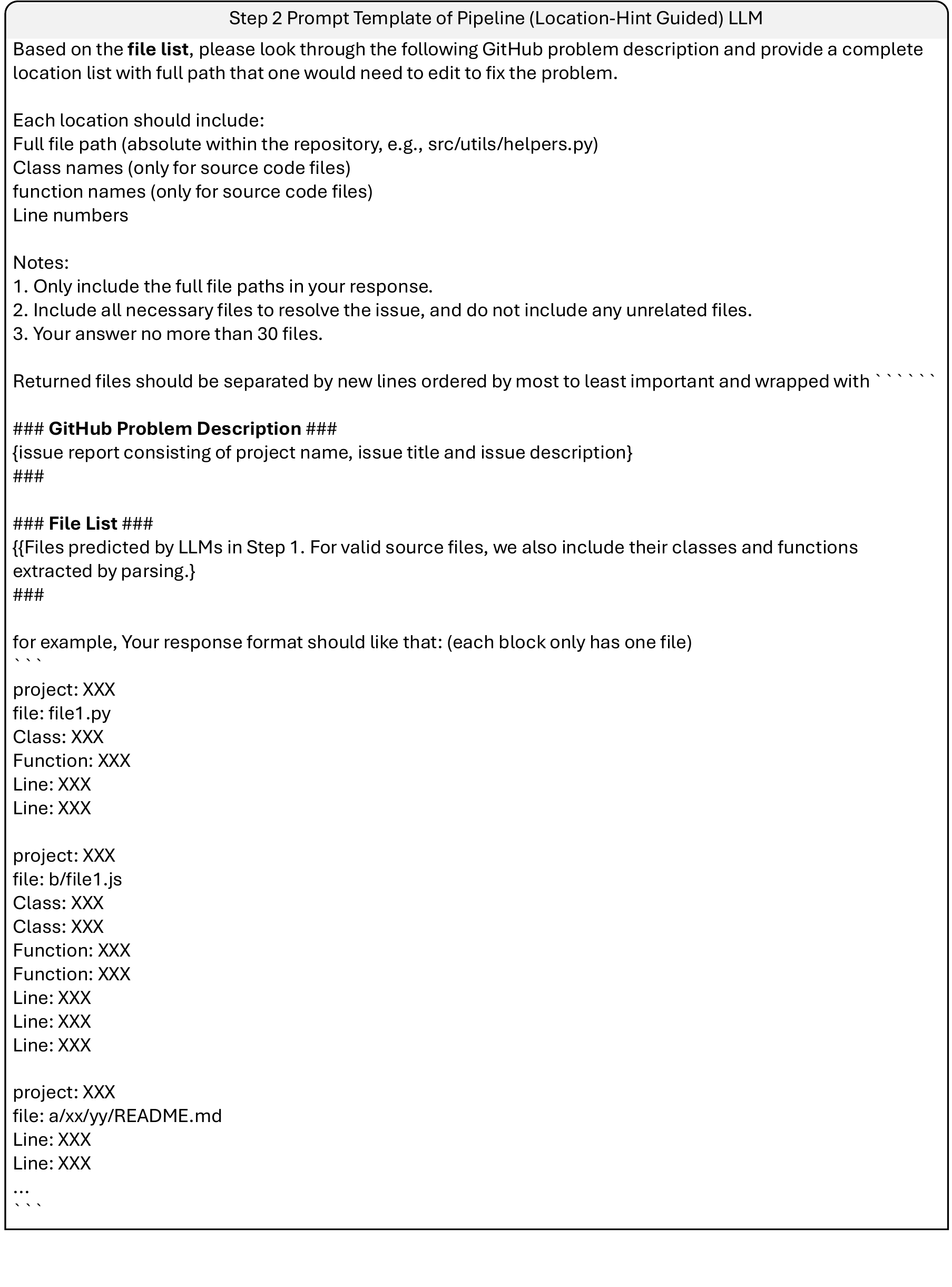}%
    \caption{Step 2 Prompt template of PipeLine (Location-Hint Guided) LLM in project localization for issue resolution. }  
    \label{fig:prompt_template_pipeline2_step2} 
    \vspace{-0.4cm}
\end{figure*}

Figure~\ref{fig:prompt_template_closed-book}, Figure~\ref{fig:prompt_template_prostructure}, Figure~\ref{fig:prompt_template_lochint}, Figure~\ref{fig:prompt_template_pipeline1_step1}, 
Figure~\ref{fig:prompt_template_pipeline1_step2},
Figure~\ref{fig:prompt_template_pipeline2_step1}, Figure~\ref{fig:prompt_template_pipeline2_step2} shows the prompt templates of LLM-based approaches.

\end{document}